\documentclass[%
reprint,
superscriptaddress,
 amsmath,amssymb,
 aps,
 pre
]{revtex4-2}

\usepackage{graphicx}
\usepackage{dcolumn}
\usepackage{bm}

\usepackage{color}
\usepackage{array}
\usepackage{ulem}
\usepackage{tabularx}
\usepackage{cancel}

\usepackage{tikz,pgfplots}
\pgfplotsset{compat=1.18}
\usetikzlibrary{positioning,hobby,decorations.markings,arrows.meta,calc,fit}
\usepackage{natbib}
\usepackage{url}

\newcommand{\pk}{}
\newcommand{\pksub}{}

\newcommand{\bfm}[1]{\mathbf{#1}}
\newcommand{\bsn}[1]{\boldsymbol{#1}}

\newcommand{\bfe}{\bfm{e}}
\newcommand{\bff}{\bsn{f}}

\newcommand{\bfu}{{\bsn{u}}}

\newcommand{\bfw}{\bsn{w}}

\newcommand{\bfq}{\bfm{q}}

\newcommand{\bfy}{{\bsn{y}}}
\newcommand{\bfr}{\bfm{r}}
\newcommand{\bfv}{{\bsn{v}}}
\newcommand{\bfx}{{\bsn{x}}}
\newcommand{\bfz}{{\bsn{z}}}
\newcommand{\bfalpha}{\bsn{\alpha}}

\newcommand{\bfphi}{\bsn{\phi}}
\newcommand{\bfPhi}{\bsn{\Phi}}
\newcommand{\bfpsi}{\bsn{\psi}}
\newcommand{\bfPsi}{\bsn{\Psi}}
\newcommand{\bfLambda}{\bsn{\Lambda}}

\newcommand{\dt}{{\text d}t}

\newcommand{\dblnk}{\begin{minipage}{12pt}\fontsize{6pt}{3pt}\selectfont{$>1$\\$<p$}\end{minipage}}

\makeatletter
\newcommand{\btriangle}{\mathpalette\btriangle@\relax}
\newcommand{\btriangle@}[2]{%
  \begingroup
  \sbox\z@{$\m@th#1\blacktriangle$}%
  \makebox[\wd\z@]{%
    \raisebox{0.0ex}{%
      \resizebox{0.9\wd\z@}{1.1\ht\z@}{%
        $\m@th#1\triangle$%
      }%
    }%
  }%
  \endgroup
}
\makeatother

\begin{document}

\title{Reduced order modelling of Hopf bifurcations for the\\ Navier-Stokes equations through invariant manifolds}

\author{Alessio Colombo}
\email{alessio.colombo@polimi.it}
\affiliation{%
    Politecnico di Milano, Milan, Italy
}%

\author{Alessandra Vizzaccaro}
\affiliation{%
    University of Exeter, Exeter, UK
}%

\author{Cyril Touz\'e}
\affiliation{%
    Institute of Mechanical Sciences and Industrial Applications (IMSIA), ENSTA - CNRS - EDF; Institut Polytechnique de Paris, Palaiseau, France
}%

\author{André de F. Stabile}
\affiliation{%
    Institute of Mechanical Sciences and Industrial Applications (IMSIA), ENSTA - CNRS - EDF; Institut Polytechnique de Paris, Palaiseau, France
}%

\author{Luc Pastur}
\affiliation{%
    Laboratoire de Mécanique et de ses Interfaces (LMI), Unité de mécanique, ENSTA; Institut Polytechnique de Paris, Palaiseau, France
}%

\author{Attilio Frangi}
\affiliation{%
    Politecnico di Milano, Milan, Italy
}%

\date{\today}

\begin{abstract} 
This work introduces a parametric simulation-free reduced order model for incompressible flows undergoing a Hopf bifurcation, leveraging the parametrisation method for invariant manifolds. Unlike data-driven approaches, this method operates directly on the governing equations, eliminating the need for full-order simulations. The proposed model is computed at a single value of the bifurcation parameter yet remains valid over a range of values. The approach systematically constructs an invariant manifold and embedded dynamics, providing an accurate and efficient reduction of the original system. The ability to capture pre-critical steady states, the bifurcation point, and post-critical limit cycle oscillations is demonstrated by a strong agreement between the reduced order model and full order simulations, while achieving significant computational speed-up.
\end{abstract}

\maketitle

\section{Introduction}

Fluid flows that obey the Navier-Stokes equations are infinite-dimensional nonlinear dynamical systems. In many cases, however, the dynamics evolves on a low-dimensional manifold. Approximating this small subset has thus been an important target for deriving efficient reduced-order models (ROMs) that qualitatively and quantitatively reproduce the full system's transient and asymptotic dynamics. Over the past decades, many different reduced-order modelling approaches have been developed, which can be broadly categorised into data-driven approaches~\cite{PODturbflows,noack2003jfm,DMD2014,kutz2016book,rowley2017arfm,loiseau2018jfm,brunton2020arfm,kaszas2022prf,bruntonSINDy,sipp2016amr,deng2020jfm,schmid2022annurev} and simulation-free methods~\cite{HaragusIooss,TITI1990,Carini2015,GallairePush,Buza:NS}.

Data-driven techniques construct ROMs by systematically extracting low-dimensional representations from data, previously collected from full-order simulations or experiments. These methods rely on optimal data transformation to embed high-dimensional dynamics into a reduced subspace. Classical linear approaches, such as Proper Orthogonal Decomposition (POD)~\cite{PODturbflows} and Dynamic Mode Decomposition (DMD)~\cite{DMD2014}, identify dominant modes via spectral decomposition. The ROMs constructed by projecting the Navier-Stokes equations onto a truncated basis of the POD modes are structurally linear-quadratic \cite{stuart1958,noack2003jfm,deng2020jfm}. 
Nonlinear techniques, such as manifold learning with deep autoencoders~\cite{lusch2018nature,chen2018neuralODE} or polynomial expansions~\cite{ssmlearn}, leverage nonlinear mappings to discover intrinsic coordinates.

When studying bifurcating flows, the system's dependence on the bifurcation parameter becomes critical. Parametric ROMs are thus indispensable, as they must capture not only the local dynamics near the equilibria, but also global topological changes across the bifurcation threshold, such as the emergence of limit cycles. To construct parametric ROMs, data-driven methods employ several strategies. Spectral approaches like POD-Galerkin~\cite{sipp2016amr,noack2003jfm} and DMD~\cite{schmid2022annurev}  can achieve accurate parametric ROMs when trained on sufficiently rich datasets spanning the parameter space. To mitigate dimensionality challenges, interpolation techniques such as Grassmannian manifold interpolation~\cite{Amsallem2009} and reduced basis methods~\cite{Rozza2007RB} have been developed. Sparse regression techniques (\textit{e.g. SINDy}~\cite{bruntonSINDy} and operator inference~\cite{Peherstorfer2016}) also address the issue by attempting to reconstruct governing equations from limited data. Nonetheless, these methods still require extensive high-fidelity data for training, and their offline computational costs grow rapidly with the system's dimensionality.

In contrast, simulation-free methods operate directly on the governing equations, eliminating the need for precomputed data. Centre manifold reduction~\cite{Haken81,gucken83,HaragusIooss} provides a rigorous framework for deriving ROMs near bifurcation points, where the dynamics is governed by a slow, low-dimensional subspace. The parametrisation method for invariant manifolds~\cite{Cabre3,Haro} generalises the technique by constructing nonlinear embeddings of invariant subspaces, showing that the graph transform method used in the center manifold theorem is only one of infinitely many possible solutions or {\it styles} of parametrisation; among these, the normal form style can also be found, allowing to make a clear link with the normal form theory~\cite{Murdock}. These methods rely on a rigorous mathematical framework that is amenable to ensuring exact convergence to the full-order solution, but only within asymptotically small neighbourhoods, as a consequence of the local nature of the underlying theoretical developments.

While the parametrisation method for invariant manifolds~\cite{Haro} has been used extensively in the field of nonlinear vibrations in recent years~\cite{Haller2016,JAIN2021How,ReviewROMGEOMNL,vizza21high,li2021periodic,opreni22high,vizza2023superharm,Martin:rotation,Frangi:electromech,Pinho:shells}, its application to bifurcating flows is still rarely documented. Interestingly, it has been used in~\cite{Carini2015} with the normal form style, without referring to the name of the general method. However, an infinity of other styles could be selected, among which the graph style, which is remarkable since all the canonical proofs of the centre manifold theorem use the graph transform. In~\cite{Carini2015}, the authors use the normal form style to solve for a centre manifold problem exactly at the Hopf bifurcation point, which is original as compared to the existing developments and well-justified within the general framework provided by the parametrisation method. The method has also been used in~\cite{Buza:NS} for the Navier-Stokes equation, but without introducing a parameter dependence. In such a case, it should be mentioned that the method is not formally designed to deal with added parameter dependence, which can be treated as proposed in~\cite{vdb:centerMORE,MingwuLi2024,Stabile:follow}. Indeed, the added direction related to the parameter is neutral, which can be seen as directly conflicting with the method's core hyperbolicity requirements~\cite{Cabre1}. However, as commented in~\cite{Haro}, the parametrisation method can be used not only to demonstrate the existence and uniqueness of invariant manifolds~\cite{Cabre1,cabre2}, but also in a more flexible way as a computational technique for obtaining high-order approximations of these manifolds. Moreover, recent extensions of the parametrisation method also considers the case of a center manifold for maps~\cite{vdb:center2020}, then extended to continuous dynamical systems including a parameter dependence~\cite{vdb:centerMORE}, thus completely relaxing the potentially too stringent conditions that might otherwise restrain the use of the method.

In this work, the parametrisation method is used as an efficient algorithm which provides a computational framework for high-order approximations of bifurcating flows. The parameter dependence is treated as proposed in~\cite{MingwuLi2024,Stabile:follow}, and we show that the technique remains effective, enabling the construction of computationally efficient ROMs valid across a range of parameter values. Following developments led in nonlinear vibration theory, where it is important to find a direct computation of the ROM that is solvable from the physical space rather than from the modal space~\cite{artDNF2020,JAIN2021How,vizza21high}, we will refer to the method as DPIM for direct parametrisation of invariant manifolds, in order to stress that the technique only needs the computation of the master eigenvectors, and in turn offers a direct nonlinear mapping from the physical degrees of freedom to the reduced subspace spanned by the selected invariant manifold.

As mentioned before, the approach presented herein shares conceptual similarities with~\cite{Carini2015}. An important difference lies in the recognition of the more general framework offered by the parametrisation method. Rewriting the developments in that setting, a flexible algorithm where the user can select either the graph style, which was not studied in~\cite{Carini2015}, or the normal form style, is proposed. Such a flexible presentation also opens the door to the question of finding the best parametrisation for a given problem. Since an infinity of styles exist, this task is difficult to address and out of the scope of the present study. Nevertheless, some parameterisations might be able to offer a larger convergence range for the solutions, as shown for example in~\cite{Stoychev:failing} for a case in nonlinear vibrations.
As compared to~\cite{Carini2015}, two other important differences are worth being underlined. First, we propose to compute the ROM for a single value of the Reynolds number which is not necessarily the critical value where the Hopf bifurcation occurs, as done in~\cite{Carini2015}. In particular, the numerical results undoubtedly show that computing the ROM after the bifurcation point, and thus using the unstable manifolds of the fixed points, allows one to significantly enlarge the range of variations of the Reynolds number for which the predictions are coincident with the full-order model. Second, an \textit{a priori} error estimate is provided, enhancing the ROM's practical utility and mitigating the limitations of operating beyond the method's theoretical framework.

The article is organised as follows. Section~\ref{sec:DPIM_NSv2} outlines the problem setting and the proposed method at a high-level. Expanding on this, Section~\ref{sec:detailed_method} provides a technical derivation of the reduced-order model. Section~\ref{sec:testcase} gives a brief description of the selected testcase, while the numerical results obtained are presented in Section~\ref{sec:numresults}. Finally, conclusions are drawn in Section~\ref{sec:conclusions}.

\section{Invariant manifold reduction for bifurcating flows}
\label{sec:DPIM_NSv2}

In this section, an overview of the proposed parametrisation method for bifurcating flows described by the Navier-Stokes equations is presented. The fluid obeys the dimensionless incompressible Navier-Stokes equations:
\begin{equation}
\label{eq:NS}
\begin{aligned}
\partial_t \bfu + \nabla \cdot (\bfu \otimes \bfu) & = \frac{1}{Re}\triangle \bfu - \nabla p,&\text { in } \Omega_t \times[0, T],\\
    0 &= \nabla \cdot \bfu  &\text { in } \Omega_t \times[0, T].
\end{aligned}
\end{equation}
where $p(\bfx, t)$ and $\bfu (\bfx, t)$ are the pressure and velocity fields, respectively, while $\bfx \in \Omega_t$ represents the space variable and $t\in [0,T]$ the time. The problem considered is such that a steady solution loses its stability for a critical value $Re_c$ of the Reynolds number through a Hopf bifurcation, which is a typical situation encountered in many examples, such as the flow past a cylinder~\cite{zdravkovich1997}. The linear stability analysis around the steady solution shows that among all the eigenvalues, a single complex conjugate pair crosses the imaginary axis at $Re = Re_c$. This pair of unstable eigenvalues will be referred to as $(\lambda, \bar{\lambda})$.

\begin{figure}[ht!]
\centering
\includegraphics[width=\linewidth]{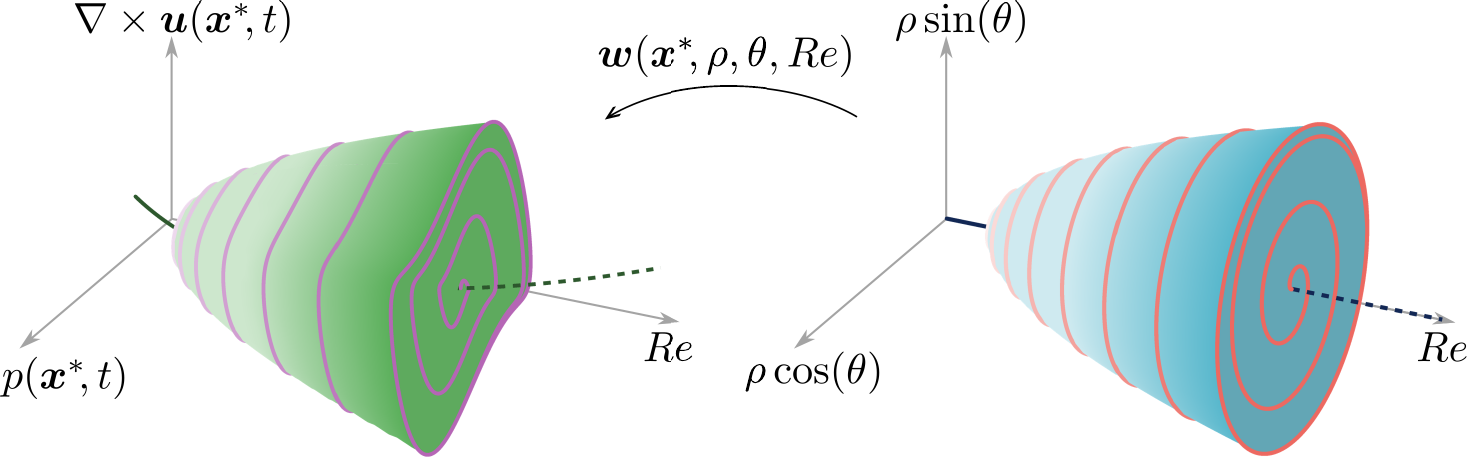}
\caption{Graphical representation of the invariant manifold in physical space for a specific point $\bfx^*$ (left), and its latent space counterpart (right). The stable (\textcolor[RGB]{45.9, 86.7, 45.9}{\rule[0.5ex]{0.2cm}{1pt}}) and unstable (\textcolor[RGB]{45.9, 86.7, 45.9}{\rule[0.5ex]{0.09cm}{1pt}\,\rule[0.5ex]{0.09cm}{1pt}}) steady solutions, and some trajectories (\textcolor[RGB]{181.05, 102., 181.05}{\rule[0.5ex]{0.2cm}{1pt}}) are reported. }
\label{fig:manifold_qualitative}
\end{figure}
The goal is to compute a single reduced-order model (ROM) valid for a range of Reynolds numbers around a user-selected value $Re_0$.
This $Re_0$ corresponds to a steady solution $(\bfu_0,p_0)$ which can be either stable (if $Re_0 < Re_c$), neutral ($Re_0= Re_c$) or unstable ($Re_0 > Re_c$). Since the proposed ROM is parametric in $Re$, it can compute both the steady and unsteady solutions corresponding to vortex shedding after the bifurcation.

In particular, given a low-dimensional invariant manifold embedded in the original system's phase space, the core idea of the method is to construct: (i) a nonlinear function that maps a set of manifold coordinates (often referred to as \textit{latent} or \textit{reduced} coordinates) to the physical ones and (ii) a reduced-order model, as an ODE in the manifold coordinates, describing the dynamics of the system on the manifold.
Given that the pair of eigenvalues losing stability through the Hopf bifurcation is the main driver of the dynamics, the proposed ROM can capture the essential features of the full-order problem with the fewest possible coordinates: two for reproducing the oscillatory behaviour and one for the $Re$-dependence. 
The proposed ROM is constructed by first computing the steady solution ($\bfu_0$,$p_0$) at a user-selected value $Re_0$, and then by identifying the unstable (or least stable if $Re_0<Re_c$) pair of eigenvalues and eigenmodes of the problem around that solution. This complex conjugate pair of eigenmodes are commonly referred to as the \textit{master} modes, as they are the main driver of the slow dynamics. Note that the selected $Re_0$ must be sufficiently close to $Re_c$, to allow for this identification.
Then, we seek to parametrise the solution manifold of the system with three coordinates: $\rho(t)$ representing the latent amplitude of oscillations, $\theta(t)$ their phase, and $Re$ the bifurcation parameter. A graphical representation is given in Fig.~\ref{fig:manifold_qualitative}. To do so, the method constructs nonlinear functions ($\bfw_\bfu,w_p$) that map the manifold coordinates to the physical fields, such that
\begin{subequations}\label{eq:maps}
    \begin{align}
    &\bfu(\bfx,t) \approx 
    \bfu_0(\bfx)+
    \bfw_\bfu(\bfx,\rho(t),\theta(t),Re),
    \\
    &p(\bfx,t) \approx 
    p_0(\bfx) +
    w_p(\bfx,\rho(t),\theta(t),Re).
\end{align}\end{subequations}
Simultaneously, it also constructs a reduced-order system of equations governing the dynamics of the manifold variables. Using the simplest formulation for this reduced dynamics yields:
\begin{subequations}\label{eq:ROMredyn}
\begin{align}
    \label{eq:ROM1}
    &\partial_t\rho = c_1(Re) \rho + 
    c_3(Re) \rho^3 + 
    c_5(Re) \rho^5 + \dots,
    \\
    \label{eq:ROM2}
    &\partial_t\theta = 
    c_0(Re) +
    c_2(Re) \rho^2 +
    c_4(Re) \rho^4 + \dots,
\end{align}
\end{subequations}
which needs to be compounded with the trivial expression for the evolution of $Re$: $\partial_t Re = 0$. Here it is important to emphasise that this specific form of the reduced dynamics is dependent on the \textit{style} of parametrisation chosen~\cite{Haro}, and Eqs.~\eqref{eq:ROMredyn} have this simple form only when the \textit{normal form style} is selected. Indeed, in that case, only the (near-)resonant monomials stay in the reduced dynamics, in line with the normal form theory~\cite{Murdock}. Given that close to the instability, the real parts of the eigenvalues are small, one can assume that the eigenspectrum is dominated by the purely complex conjugate imaginary parts, such that only the resonant monomials corresponding to Eq.~\eqref{eq:ROMredyn} will be computed by the DPIM~\cite{TouzeCISM,TouzeCISM2}. However, different styles of parametrisation can be selected to compute the reduced dynamics and nonlinear mappings. This point will be further addressed in the remainder of the paper.


The unknown quantities are obtained by enforcing two key properties: the tangency of $(\bfw_\bfu,w_p)$ to the master eigenmodes at $Re_0$, which ensures that the resulting manifold is the one corresponding to the dynamics of interest; the invariance of the manifold under the action of the original dynamical system. The latter is imposed by requiring that Eqs.~\eqref{eq:maps} and~\eqref{eq:ROMredyn} satisfy Eq.~\eqref{eq:NS}.

The explicit expressions for $(\bfw_\bfu,w_p)$ as a function of $(\rho,\theta,Re)$, and the scalars $c_k$ as a function of $Re$ are all constructed as arbitrary order polynomials in the variables ($z_1,z_2,z_3)=(\rho e^{+i\theta}$, $\rho e^{-i\theta}$, $\frac{1}{Re}-\frac{1}{Re_0}$), as this choice allows for algorithmic simplifications. Due to the polynomial form of these expressions, the algorithm only requires repeated solutions of linear systems, each of the size of the original Eq.~(\ref{eq:NS}) and as many as the number of monomials. A more detailed derivation of the procedure is provided in the next section.
The output of the method is thus the coefficients of the polynomial expansion of $(\bfw_{\bfu},w_p)$ as well as those of $c_j(Re)$, up to a desired order. The full fields at a given $Re$ can be reconstructed by first solving the reduced order system of Eqs.~\eqref{eq:ROMredyn}, and then using the resulting $\rho (t),\theta (t)$ in Eqs.~\eqref{eq:maps}. Note that the dependence of $\bfu$ and $p$ with respect to time is only through the manifold coordinates, meaning that the dynamics of the ROM solution always lies on the manifold.

Referring to the reduced-order system, Eqs.~\eqref{eq:ROMredyn}, it can be seen that the ROM provides explicit expressions for the instantaneous rate of decay (or amplification) of the oscillations $\partial_t\rho$, and their instantaneous frequency $\partial_t\theta$, as a direct consequence of using the normal form style in the parametrisation procedure. In such a case, the technique, when truncated to the third order, directly retrieves the Stuart-Landau equations used \textit{e.g.} in~\cite{SIPP_LEBEDEV_2007,GallairePush}, which are thus automatically computed from the sole geometrical input (mesh and boundary conditions). It also provides higher-order approximations to these equations, depending on the user's choices. If another style of parametrisation is selected, as more thoroughly explained in the next section, then the reduced dynamics has a more complex formulation.


The output of the ROM can be exploited to obtain several quantities of interest. First, any other steady solution $(\bfu_s,p_s)$ at a given value $Re\neq Re_0$ can be retrieved, by simply setting $\rho=0$ in Eqs.~\eqref{eq:maps}. The slow eigenvalues $(\lambda,\bar{\lambda})$ governing the stability of the steady solution can also be readily obtained as a function of $Re$ from the reduced order system of Eqs.~\eqref{eq:ROMredyn} in the limit $\rho\rightarrow 0$:
\begin{equation}
\label{eq:eigen_at_Re}
    \lambda(Re) = c_1(Re) + i\, c_0(Re).
\end{equation}
This expression provides a means to predict the Hopf bifurcation point, which corresponds to the critical Reynolds number $Re_c$ such that $c_1 (Re_c) = 0$.

The limit cycles occurring at $Re>Re_c$ can also be found directly by imposing $\partial_t\rho = 0$ in Eq.~\eqref{eq:ROM1}, and looking for a non-trivial solution $\rho_{lc}>0$. From the identified amplitude $\rho_{lc}$ of the limit cycle, the oscillation frequency can be computed from Eq.~\eqref{eq:ROM2} and the full fields reconstructed from Eqs.~\eqref{eq:maps}.

Lastly, if one is interested in the slow transient dynamics, starting from an initial state in the vicinity of a steady solution $(\bfu_s, p_s)$ and evolving towards the limit cycle, the time integration of Eq.~\eqref{eq:ROM1} is required. However, since this is a single degree of freedom differential equation, its time integration has a minimal computational cost.

\section{Algorithmic details: derivation of the parameter-dependent ROM for the Navier-Stokes equations}
\label{sec:detailed_method}

This section aims to provide more insights into the detailed derivation of the ROM for the Navier-Stokes equations. The presentation of the method relies on the implementation of the DPIM as detailed in~\cite{vizza2023superharm,Stabile:follow}. To fit the general framework presented in~\cite{vizza2023superharm,Stabile:follow}, a recast of the Navier-Stokes equations is needed, which is here detailed.

First, the problem is rewritten around the steady solution ($\bfu_0,p_0$) corresponding to a selected $Re_0$. To this end, the following variables are introduced:
\begin{equation}
\begin{aligned}
    &\eta_0 = 1/Re_0, \quad \eta = 1/Re,\\
    &\eta' = \eta- \eta_0,\\
    & \bfu' = \bfu - \bfu_0,\\
    & p' = p - p_0.
\end{aligned}
\end{equation}
The Navier-Stokes equations around the steady solution can thus be rewritten as:
\begin{subequations}\label{eq:NS-0}
\begin{align}
&
\partial_t \bfu' + 
\mathcal{L}_0(\bfu') +
\nabla \cdot (\bfu' \otimes \bfu') 
= 
\eta' \triangle \bfu_0 +
\eta' \triangle \bfu' -
\nabla p', \label{eq:NS-0a}
\\
&
\nabla \cdot \bfu' = 0, \label{eq:NS-0b}
\\
&
\partial_t \eta' = 0. \label{eq:NS-0c}
\end{align}
\end{subequations}
In particular, the last equation has been added in order to define the bifurcation parameter as a variable of the dynamical system, which is needed to embed the parameter dependence in the ROM. Note that the choice of introducing the parameter-dependence through $\eta'$ rather than $Re$ is the same as in~\cite{GallairePush}, where it was found to be optimal. Moreover, the linear operator $\mathcal{L}_0$ has been introduced for compactness and reads
\begin{equation}
    \mathcal{L}_0(\bfu')
    =
    \nabla \cdot 
    (\bfu' \otimes \bfu_0 +
    \bfu_0 \otimes \bfu' ) -
    \eta_0 \triangle \bfu'
\end{equation}
Lastly, to obtain the required form, the unknowns are grouped into a single vector as:
\begin{equation}
    \bfy = \begin{bmatrix}
        \bfu'\\
        p'\\
        \eta'
    \end{bmatrix},
\end{equation}
and the following operators are introduced:
\begin{subequations}
\label{eq:operators}
    \begin{align}
    \mathcal{B}\left(
    \bfy
    \right)
    &=
    \begin{bmatrix}
        \bfu'\\
        0\\
        \eta'
    \end{bmatrix},\\
    \mathcal{A}\left(
    \bfy
    \right)
    &=
    \begin{bmatrix}
    -\mathcal{L}_0(\bfu')-\nabla p'+\Delta\bfu_0 \eta'\\
    -\nabla \cdot \bfu'\\
    0
    \end{bmatrix},\\
    \mathcal{Q}\left(
    \bfy
    \right)
    &=
    \begin{bmatrix}
    -\nabla \cdot (\bfu' \otimes \bfu') +
    \eta' \Delta \bfu' \\
    0\\
    0
    \end{bmatrix}.
    \end{align}
\end{subequations}

The original Navier-Stokes equations can be finally rewritten as
\begin{equation}
\label{eq:strong_DAE}
    \mathcal{B}(\partial_t \bfy) = 
    \mathcal{A}(\bfy) + \mathcal{Q}(\bfy).
\end{equation}
This is the general form treated in \cite{vizza2023superharm,Stabile:follow}, with the non-linearities embedded entirely in the quadratic operator~$\mathcal{Q}$. Note that any partial differential equations (PDE) with smooth non-linearities can be written in this form thanks to the so-called \textit{quadratic recast}~\cite{Guillot:recast}. In particular, the discretised weak formulation of Eq.~\eqref{eq:strong_DAE} leads to a quadratic differential-algebraic system of equations (DAE) due to the incompressibility constraint, which is precisely the framework used in~\cite{vizza2023superharm,Stabile:follow}.

Starting from Eq.~\eqref{eq:strong_DAE}, the method seeks to parametrise an invariant manifold by constructing a nonlinear function $\bfw_\bfy$ mapping the manifold coordinates $\bfz$ to the original field $\bfy$, and a reduced-order ODE system, with vector field $\bff$, in the manifold coordinates governing the reduced dynamics. The time evolution of the original system is then approximated by the time evolution on the manifold:
\begin{subequations}
\label{eq:map_and_dyn}
\begin{align}
    &\bfy(\bfx, t) \approx \bfw_\bfy(\bfx,\bfz(t)),\\
    &\partial_t \bfz(t)=\bff(\bfz(t)).
\end{align}
\end{subequations}
The so-called \textit{invariance equation}~\cite{Haro}, through which the invariance property of the resulting manifold is imposed, can now be explicitly obtained by substituting Eqs.~\eqref{eq:map_and_dyn} in Eq.~\eqref{eq:strong_DAE}, and reads
\begin{align}
\label{eq:invariance}
    \mathcal{B} 
    (\nabla_\bfz \bfw_\bfy \bff )
    = 
    \mathcal{A}(\bfw_\bfy) + \mathcal{Q}(\bfw_\bfy).
\end{align}
Note that the time-dependence of the system is only through the manifold coordinates $\bfz$, therefore, time does not explicitly appear in the equation.

The proposed algorithm constructs the expressions for $\bfw_\bfy$ and $\bff$ as polynomials in $\bfz$, as done in the parametrisation method. Given an order of expansion $o$, defining $m_p$ as the number of monomials in $\bfz$ of order $p$, and adopting the multi-index notation, these expressions read:
\begin{subequations}
\label{eq:poly_expansion_map_and_dyn}
\begin{align}
    \bfw_\bfy(\bfx,\bfz)
    &= 
    \sum_{p=1}^o \sum_{k=1}^{m_p}
    \bfw_\bfy^{(p,k)}(\bfx) \bfz^{\bfalpha(p,k)},\\ 
    \bff(\bfz) &= 
    \sum_{p=1}^o \sum_{k=1}^{m_p}
    \bff^{(p,k)} \bfz^{\bfalpha(p,k)}.
\end{align}
\end{subequations}
In particular, $\bfz^{\bfalpha(p,k)}$ denotes the $k$-th monomial of order $p$ and $\bfalpha(p,k)$ its corresponding exponents:
\begin{align*}
    \bfz^{\bfalpha(p,k)} &= \prod_{i=1}^{|\bfz|}\bfz_i^{\alpha_i(p,k)},\\
    \bfalpha(p,k) &= \{\alpha_1(p,k), \ldots, \alpha_{|\bfz|}(p,k)\},
\end{align*}
where $|\bfz|$ is the dimension of the manifold, that can be set arbitrarily, as shown below.

The unknown fields $\bfw_\bfy^{(p,k)}$ and coefficients $\bff^{(p,k)}$ are computed by substituting Eqs.~\eqref{eq:poly_expansion_map_and_dyn} in the invariance equation, obtaining, for each monomial, a so-called \textit{homological equation}.
We address the order 1 homological first, then orders $p \geq 2$.

\subsection{Order 1}\label{subsec:order1_homol}

The invariance property, if the manifold contains the origin, implies its tangency to a set of modes at the origin. This set, however, can be selected arbitrarily, thereby determining both the dynamics captured by the manifold and its dimension $|\bfz|$.
In particular, let $\bfphi$ and $\bfLambda$ be the collections of unknown fields $\bfw_\bfy^{(1,k)}$ and coefficients $\bff^{(1,k)}$ of order 1:
\begin{align}
\bfphi &= [\bfw_\bfy^{(1,1)}, \ldots, \bfw_\bfy^{(1,|\bfz|)}],\\
\label{eq:Lambda}
\bfLambda &=[\bff^{(1,1)},\ldots,\bff^{(1,|\bfz|)}],
\end{align}
as such notation will soon prove to be natural. With this choice, the set of homological equations of order 1 can be written compactly:
\begin{equation}
\label{eq:order_1_homol}
    \mathcal{B} (\bfphi \cdot \bfLambda_{j}) = \mathcal{A} (\bfphi_j)\, \qquad \forall j = 1,\ldots,|\bfz|.
\end{equation}
One can recognise here the same structure as the linear eigenproblem. Importantly, since $\bfphi$ and $\bfLambda$ remain to be specified, any subset of eigenmodes and their associated eigenvalues yields a valid solution. This choice determines both the size of the manifold $|\bfz|$ and the dynamics it can capture.

Note that choosing as columns of $\bfphi$ linear combinations of the eigenmodes also satisfies Eq.~\eqref{eq:order_1_homol}; however, selecting exactly the eigenmodes is the most computationally efficient choice, as it diagonalises $\bfLambda$ and ensures the decoupling of all homological equations of the same order, as previously mentioned. 

Eqs.~\eqref{eq:poly_expansion_map_and_dyn} can finally be rewritten to make the tangency explicit:
\begin{subequations}
\begin{align}
    \bfw_\bfy(\bfx,\bfz) &= \bfphi(\bfx) \bfz \,+\, \sum_{p=2}^o\sum_{k=1}^{m_p}\bfw_\bfy^{(p,k)}(\bfx)\bfz^{\bfalpha(p,k)},\\
    \bff(\bfz) &= \bfLambda\bfz \,+\, \sum_{p=2}^o\sum_{k=1}^{m_p}\bff^{(p,k)}\bfz^{\bfalpha(p,k)}.
\end{align}
\end{subequations}

\subsection{Order $p \geq 2$}\label{subsec:orderp_homol}
The general form of the order-$p$ homological equation is:
\begin{equation}
\label{eq:order_p_homological_discrete}
    \mathcal{S}^{(p,k)} \left(\bfw_\bfy^{(p,k)}\right) + \sum_{s=1}^{|\bfz|} \mathcal{B} \left( \bfphi_s f_s^{(p,k)} \right) = \bfq^{(p,k)},
\end{equation}
where $\mathcal{S}^{(p,k)}=\sigma^{(p,k)} \mathcal{B} - \mathcal{A}$, and $\sigma^{(p,k)} = \sum_{s=1}^{|\bfz|} \alpha_s(p,k) \lambda_s$. The full derivation is omitted in the continuous case, but is detailed in the discretised case in Appendix~\ref{sec:discrete}, for the interested reader.
The term $\bfq^{(p,k)}$ collects lower-order contributions, making the homological equations decoupled and solvable monomial-by-monomial. We therefore drop the notation $(p,k)$ in the remainder of the section, to promote a lighter presentation. However, each equation remains under-determined due to the simultaneous appearance of both the nonlinear mapping coefficients $\bfw_\bfy\pk$ and the reduced dynamics coefficients $f_s\pk$.
This is a classical feature of the parametrisation method~\cite{Haro}, which states that an infinity of solutions to these equations exist. Each of these solutions is called {\it a style of parametrisation}, and corresponds to a particular selection of the variables that will be used to describe the invariant manifold and its resulting dynamics.

To better explain how the different styles are selected in the solution procedure, the adjoint eigenmodes $\bfpsi$ are introduced:
\begin{equation}
\label{eq:adjoint_eigenproblem}
    \mathcal{B}^* (\bfpsi \cdot \bfLambda_{j}) = \mathcal{A}^* (\bfpsi_j), \qquad \forall j = 1,\ldots,|\bfz|.
\end{equation}
where $\mathcal{B}^*$ and $\mathcal{A}^*$ denote the adjoint operators of $\mathcal{B}$ and $\mathcal{A}$, defined with respect to a suitable inner product $\langle \cdot, \cdot \rangle$.
The adjoint eigenmodes satisfy the biorthogonality relationships:
\begin{subequations}
\label{eq:biorthogonality}
\begin{align}
    \langle \bfpsi_i, \mathcal{B} (\bfphi_j) \rangle &= \delta_{ij},\\
    \langle \bfpsi_i, \mathcal{A}(\bfphi_j) \rangle &=  \lambda_j \delta_{ij},
\end{align}
\end{subequations}
where $\delta_{ij}$ is the Kronecker delta. Exploiting Eqs.~\eqref{eq:adjoint_eigenproblem}-\eqref{eq:biorthogonality}, the following relation also holds:
\begin{equation}
    \langle \bfpsi_j, \mathcal{A}(\, \cdot \,) \rangle = \lambda_j \langle \bfpsi_j, \mathcal{B} (\, \cdot \,) \rangle.
\end{equation}
Projecting Eq.~\eqref{eq:order_p_homological_discrete} onto the $j$-th adjoint eigenmode allows deriving a scalar equation for an arbitrary monomial of a given order. It is then straight-forward to analyse the solutions and remove the underdeterminacy with the different styles. The projection yields:
\begin{equation}
\label{eq:projected_homological}
    (\sigma\pk - \lambda_j) \langle \bfpsi_j, \mathcal{B} (\bfw_{\bfy}\pk) \rangle + f_j\pk = \langle \bfpsi_j, \bfq\pk \rangle.
\end{equation} 
Eq.~\eqref{eq:projected_homological} contains two unknowns: $\langle \bfpsi_j, \mathcal{B} (\bfw_{\bfy}\pk) \rangle$, which is the projection of the nonlinear mapping terms $\bfw_{\bfy}$, and the reduced dynamics coefficients $f_j\pk$; consequently, an infinite number of possible solutions exist~\cite{Haro}. Nevertheless, two opposite choices are specifically meaningful and worth being detailed. A first solution consists of systematically enforcing $\langle \bfpsi_j, \mathcal{B} (\bfw_{\bfy}\pk) \rangle = 0$. By doing so, the simplest formulation for the nonlinear mappings is found, as this choice is equivalent to defining a graph transform between the master modal coordinates and the slave ones~\cite{Haro,TouzeCISM2}. Consequently, this choice is referred to as the {\it graph style}.

In contrast, a second strategy involves selecting the simplest reduced-order dynamics, by systematically enforcing $f_j\pk=0$ whenever it is possible.
However, when $\sigma\pk = \lambda_j$, Eq.~\eqref{eq:projected_homological} reduces to $f_j\pk = \langle \bfpsi_j, \bfq\pk \rangle$, such that the choice $f_j\pk=0$ is not possible. This case corresponds to a nonlinear resonance~\cite{Murdock}, a common feature of the normal form procedure. The choice $f_j\pk=0$ is therefore possible only for the non-resonant monomials, leading to a partial normal form over the master coordinates only, thus giving the name {\it normal form style}.

In a numerical and automated context, care must be taken with the fact that a near-resonance scenario might occur, when the resonance condition is not exactly verified but only approximated. In such a near-resonance case, with $0 < |\sigma\pk - \lambda_j| \ll 1$, setting $f_j\pk = 0$ would introduce numerical stiffness and discontinuities as exact resonance is approached. We thus set $\langle \bfpsi_j, \mathcal{B} (\bfw_{\bfy}\pk) \rangle = 0$ in this case too, to ensure smooth dependence on the parameters.

To provide an efficient treatment of both the resonance conditions and the choice of the parametrisation style, it is convenient to introduce the (near-)resonant set for each considered monomial~\cite{vizza2023superharm}, defined as:
\begin{equation}
\label{eq:resonant_set}
    \mathcal{R}\pksub = \{ j \in \, [1,|\bfz| \,] \ | \ \sigma\pk \approx \lambda_j \}.
\end{equation}
For $j \in \mathcal{R}\pksub$, the constraint $\langle \bfpsi_j, \mathcal{B} (\bfw_{\bfy}\pk) \rangle = 0$ is enforced, as this consistently handles both exact and near-resonance. Conversely, when $j \notin \mathcal{R}\pksub$, either the projection of $\bfw_\bfy\pk$ or the reduced dynamics $f_j\pk$ may be set to zero.

The two main parametrisation styles can thus be easily formulated as follows in the algorithmic procedure. The graph style is selected by enforcing the conditions:
\begin{equation}
 \label{eq:constraints_graph_style}
     \langle \bfpsi_j, \mathcal{B} (\bfw_{\bfy}\pk) \rangle = 0, \qquad \forall j = 1,\ldots,|\bfz|.
 \end{equation}

On the other hand, the normal form style is obtained when selecting:
\begin{subequations}
 \label{eq:constraints_normal_form_style}
 \begin{align}
     \langle \bfpsi_j, \mathcal{B} (\bfw_{\bfy}\pk) \rangle = 0, \qquad &j \in \mathcal{R}\pksub,\\
     f_j\pk = 0, \qquad &j \notin \mathcal{R}\pksub.
 \end{align}
 \end{subequations}

All the other possible solutions or styles of parametrisation are called {\it mixed styles}~\cite{Haro}. They consist of mixing the different possible choices (graph or normal form styles) depending on the monomial. From an algorithmic point of view, all the different styles can be easily implemented based on a simple rule derived from the resonance sets~$\mathcal{R}$. In the remainder of the study, only the graph style and the normal form style will be tested in the numerical examples.

\section{Testcase}
\label{sec:testcase}

\begin{figure}[ht]
    \includegraphics[width=\linewidth]{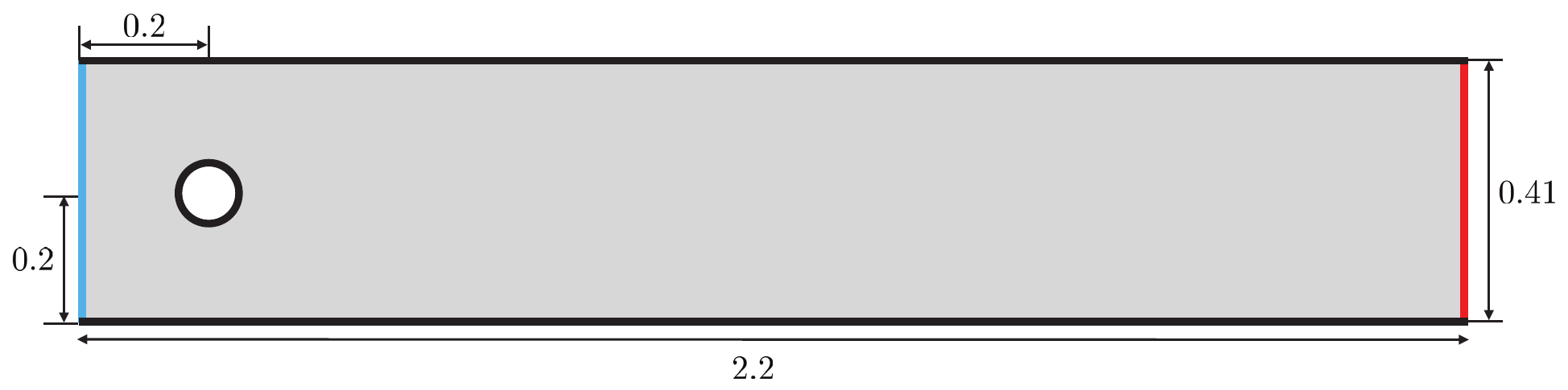}
    \caption{Geometry of the Turek-Sch\"afer benchmark for the flow around a cylinder in a 2D channel. On the boundary, no-slip (\textcolor[HTML]{000000}{\rule[0.5ex]{0.2cm}{1pt}}), non-homogeneous Dirichlet (\textcolor[HTML]{00b0f0}{\rule[0.5ex]{0.2cm}{1pt}}), and homogeneous Neumann (\textcolor[HTML]{ff0000}{\rule[0.5ex]{0.2cm}{1pt}}) boundary conditions are applied.}
    \label{fig:turek-schafer_geom}
\end{figure}

The testcase considered in this work is the Turek-Sch\"afer benchmark for the flow around a cylinder in a 2D channel~\cite{Schafer1996} (Fig.~\ref{fig:turek-schafer_geom}). The circular cylinder problem is widely studied for its role in illustrating supercritical Hopf bifurcations, where fluid instabilities lead to periodic vortex shedding. Note that the proposed algorithmic procedure can treat any flow exhibiting a Hopf bifurcation, although the specific test case affects the range within which convergence occurs.

The discretised system counts 17973 degrees of freedom. The channel is $2.2$ units long and $0.41$ units wide, with the cylinder centred at $(0.2,0.2)$ and having a diameter $D$ of $0.1$ units. The Reynolds number is defined as $Re = \bar{U} D/\nu$, with $\bar{U}$ the mean inlet velocity and $\nu $ the kinematic viscosity of the fluid. 
The inflow condition is a Poiseuille flow profile:
\begin{equation*}
    \bfu(0,y) = \frac{6 \, y \, (H-y)}{H^2}\bfe_x,
\end{equation*}
where $\bfe_x$ is the unit vector in the $x$-direction, $H$ is the width of the channel, and the scaling factor is chosen to ensure that the mean inflow velocity $\bar{U}$ equals 1. The velocity field satisfies the no-slip condition $\bfu=\bf0$ on the cylinder and at the top and bottom boundaries. A no-stress condition is applied at the outlet. For the selected set of parameters, a Hopf bifurcation occurs at $Re_c \approx 49.03$: at this point, the steady-state solution becomes unstable, giving rise to oscillations and vortex shedding in the wake of the cylinder.

\section{Numerical results}
\label{sec:numresults}
Following the proposed algorithmic procedure, we now present the numerical results obtained on the selected testcase. Unless otherwise stated, all the ROMs presented in this section have been computed with an order 5 polynomial expansion in the manifold variables and using normal form style~\eqref{eq:constraints_normal_form_style}, to balance between computational efficiency and precision of the resulting model. The results are compared with direct numerical simulations performed by advancing the discretised system using a Crank-Nicholson scheme, and the spatial nonlinearity is handled by performing Newton sub-iterations until convergence is reached.

As discussed in Section~\ref{sec:DPIM_NSv2}, all the computed ROMs will have a minimal dimension of 3, with two normal coordinates corresponding to the master mode, and one coordinate for the parameter with neutral dynamics. Since this equation is trivial, the parameter dependence can easily be replaced in the reduced dynamics, such that the ROM has an effective dimension of 2. The solution to the linear problem and the eigenvalues trajectory are detailed in Appendix~\ref{appendix:C}.

\begin{figure}[ht]
    \includegraphics[width=\columnwidth]{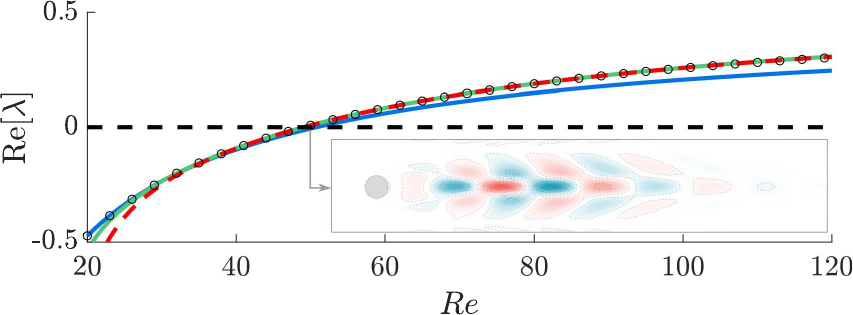}
    \includegraphics[width=\columnwidth]{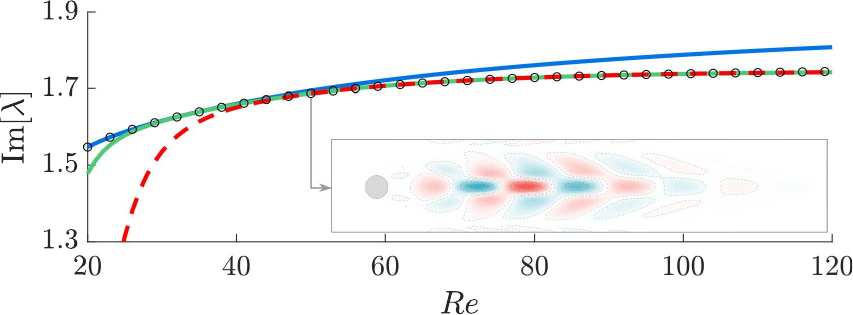}
    \caption{Real (top) and imaginary (bottom) parts of the eigenvalue of the bifurcating mode over a range of $Re$. The values computed by the FOM ($\circ$), requiring a linear eigenproblem per point, are well aligned with those predicted \textit{a priori} by three ROMs at $Re_0 = 20$ (\textcolor[HTML]{0073e6}{\rule[0.5ex]{0.2cm}{1pt}}), $Re_0 = Re_{c}$ (\textcolor[HTML]{50c878}{\rule[0.5ex]{0.2cm}{1pt}}) (corresponding to the choice made in~\cite{Carini2015}), and $Re_0 = 70$~(\textcolor[HTML]{ff0000}{\rule[0.5ex]{0.2cm}{1pt}}).
    The real and imaginary parts of the corresponding eigenmode at bifurcation are also shown. Being the eigenmode defined up to an arbitrary phase, their allocation to the top and bottom plots is only for illustrative purposes.}
    \label{fig:eigenvalue_prediction}
\end{figure}

As a first outcome, the prediction of the Hopf bifurcation point is investigated for different choices of $Re_0$, \textit{i.e.} the user-selected value for constructing the ROM.
A first parametrisation is performed for a value of the Reynolds number far from the critical Reynolds, specifically $Re_0 = 20$, and the variation of the least stable eigenvalue pair computed by the ROM is tracked as $Re$ increases, using Eq.~\eqref{eq:eigen_at_Re}. The model precisely predicts the Reynolds number at which the Hopf bifurcation occurs, as shown in Fig.~\ref{fig:eigenvalue_prediction}. In particular, the variation of the eigenvalue with $Re$ agrees with the reference full-order solution for a large interval, with the computed bifurcation point deviating from the actual value by only 3.5\%. Two more ROMs, constructed with $Re_0 = Re_{c}$ and $Re_0 = 70$, are also reported, showing an even better predictive capability to both recover the bifurcation point and capture the eigenvalues' behaviour for $Re$ values after the critical Reynolds. Notably, the model constructed with $Re_0 = Re_{c}$ coincides with that of~\cite{Carini2015}, since this choice of $Re_0$ coupled with normal form style inherently reproduces their centre manifold-based solution strategy.

Beyond eigenvalue tracking, leveraging Eqs.~\eqref{eq:maps} to recover the linear part of the maps also enables the reconstruction of the corresponding eigenmode. In Fig.~\ref{fig:eigenvalue_prediction}, we present the real and imaginary parts of the ROM eigenmode shape for the vorticity at $Re = Re_c$. These are indistinguishable from the full-order model (FOM) eigenmodes, which are thus not reported.

\begin{figure}[thb]
    \includegraphics[width=\linewidth]{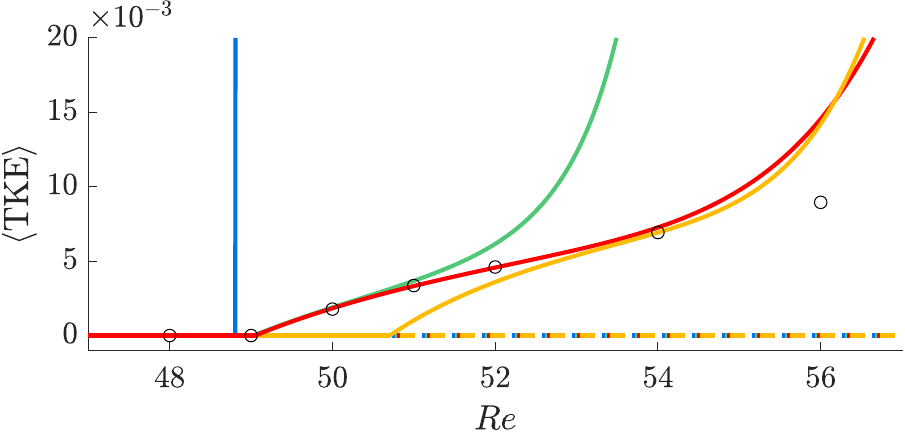}
    \caption{Bifurcation diagram of the average turbulent kinetic energy with respect to the Reynolds number: FOM ($\circ$); ROMs at $Re_0 = 20$~(\textcolor[HTML]{0073e6}{\rule[0.5ex]{0.2cm}{1pt}}), $Re_0 = Re_{c}$~(\textcolor[HTML]{50c878}{\rule[0.5ex]{0.2cm}{1pt}}) \cite{Carini2015}, $Re_0 = 70$~(\textcolor[HTML]{ff0000}{\rule[0.5ex]{0.2cm}{1pt}}), and $Re_0 = 80$~(\textcolor[HTML]{ffbb00}{\rule[0.5ex]{0.2cm}{1pt}}). The expansion point has a large effect on the accuracy of the prediction, both in the vicinity of the bifurcation and at high $Re$.}
    \label{fig:bifurcation}
\end{figure}

Let us now consider the prediction of the unsteady dynamics and the limit cycle oscillations. Due to its local nature, the method is inherently well-suited to approximate the behaviour close to the steady solution.
Conversely, the limit cycle oscillations are distant from the steady solution, which makes their prediction more challenging~\cite{Stabile:follow}. A good fit for the limit cycle, therefore, becomes a stronger indicator of the method's overall performance. To assess the effectiveness of the ROM in capturing this behaviour, we examine the bifurcation diagram, which provides a clear representation of how the amplitude of the limit cycle grows as the Reynolds number increases.
To construct this diagram, a global descriptor of the flow behaviour based on the Turbulent Kinetic Energy (TKE) is introduced, \textit{i.e.} the kinetic energy of the velocity fluctuations per unit mass.
To quantify the overall TKE, we consider its integral mean over the domain $\Omega$ and one oscillation period $T$:
\begin{equation*}
    \langle \text{TKE} \rangle = \frac{1}{|\Omega|} \frac{1}{T} \int_{\Omega}\int_0^T \frac{1}{2}\left||\bfu(\bfx,t)-\bar{\bfu}(\bfx)|\right|_2^2 \, \dt \, \text{d}\bfx.
\end{equation*}
where $\bar{\bfu}(\bfx)$ is the mean of the flow over one oscillation period in the permanent regime, commonly referred to as the \textit{mean flow}.
Fig.~\ref{fig:bifurcation} reports the $\langle \text{TKE} \rangle$ predicted by different ROMs.

When the expansion point is selected well before the instability, here illustrated with $Re_0 = 20$, even though the bifurcation point is correctly predicted, the ROM is unable to describe the limit cycles and shows a divergence, likely due to the combined effect of the large distance in parameter space and the shift in stability of the manifold.
Computing the ROM expansion at $Re_0 = Re_c$, which was the choice retained in~\cite{Carini2015}, allows recovering a correct estimate of the limit cycle's amplitudes up to $Re \approx 51$, corresponding to a shift of 4\% of the Reynolds number. This can be importantly improved by computing the ROM for a running value selected after the bifurcation point, as shown in  Fig.~\ref{fig:bifurcation} with the case $Re_0=70$. In this case, the prediction of the limit cycles is excellent in the interval $Re \in [Re_c,54]$, corresponding to a shift of 10\%. This demonstrates that the range of convergence is strictly tied to the selected expansion point for the parametrisation, and that the best results are obtained when the unstable manifold is parametrised by the method, as underlined in~\cite{MingwuLi2024,Stabile:follow} in different contexts. Moreover, this also highlights that the range of predictions given by restricting the computation to the centre manifold, as in~\cite{Carini2015}, can be significantly enlarged.

To test whether an even larger $Re_0$ could further improve the results, a final ROM around $Re_0 = 80$ is considered. Although this model outperforms the one computed at $Re_0=70$ over a small range of $Re$, outside this range the performance is overall worse, and it also fails at capturing the bifurcation. This suggests that the optimal expansion point might be closer to $Re_0=70$.


\begin{figure}[thb]
    \includegraphics[width=\linewidth]{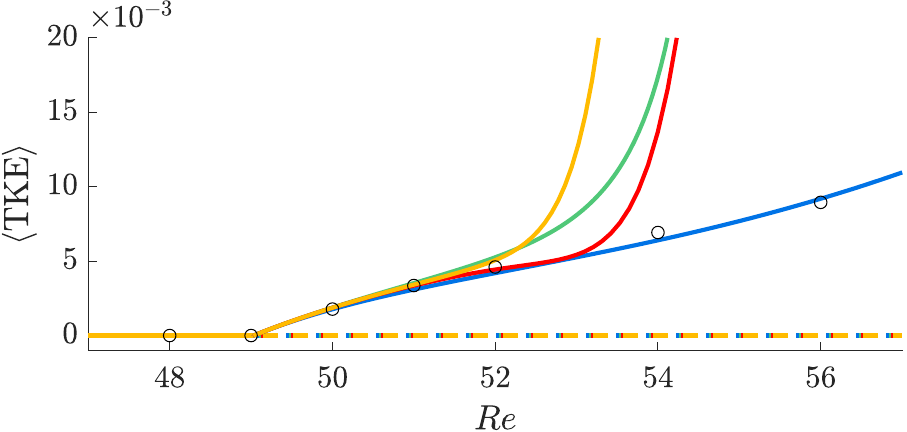}
    \caption{Bifurcation diagram of the average turbulent kinetic energy with respect to the Reynolds number: FOM ($\circ$); ROM at $Re_0 = Re_{c}$, using the graph style parametrisation~\eqref{eq:constraints_graph_style}, at orders $3$~(\textcolor[HTML]{0073e6}{\rule[0.5ex]{0.2cm}{1pt}}), $5$~(\textcolor[HTML]{50c878}{\rule[0.5ex]{0.2cm}{1pt}}), $7$~(\textcolor[HTML]{ff0000}{\rule[0.5ex]{0.2cm}{1pt}}), and $9$~(\textcolor[HTML]{ffbb00}{\rule[0.5ex]{0.2cm}{1pt}}). The ROMs agree up to $Re \approx 51$, but separate for higher $Re$ where the asymptotic expansion has not reached convergence.}
    \label{fig:bifurcation_orders}
\end{figure}

To further illustrate the capabilities of the produced ROMs, their convergence in terms of the selected order $o$ of the polynomial expansions used in Eqs.~\eqref{eq:poly_expansion_map_and_dyn}, is investigated in Fig.~\ref{fig:bifurcation_orders}. For the sake of illustration, the graph style parametrisation is now selected, following the choice given in Eq.~\eqref{eq:constraints_graph_style}, and increasing orders (3, 5, 7 and 9) are reported. The expansion point $Re_0$ is set to $Re_0=Re_c$, meaning that the centre manifold is here constructed using the graph transformation method typically used in this context, see {\it e.g.}~\cite{gucken83}. As usually reported with such asymptotic expansions~\cite{Stabile:morfesym,GROLET:high25}, increasing the order leads to higher accuracy but only within the range of convergence of the series. This is retrieved in the present case, where the prediction of the limit cycle's amplitude close to the instability is more and more accurate with increasing orders, as long as the validity limit is not reached, which is here estimated at about $Re=51$. In this specific case, the order three expansion produces a correct estimate for a large interval of Reynolds number. This result is however incidental and cannot be generalised. Finally, for this problem, we note that the graph style solution gives results which are very close to those given using normal form style. This is again incidental, and numerous cases have been reported where the choice of the style is of importance to enhance the validity range of the approximations, see {\it e.g.}~\cite{vizza21high,Stoychev:failing}.

\begin{figure*}[thb]
    \includegraphics[width=\textwidth]{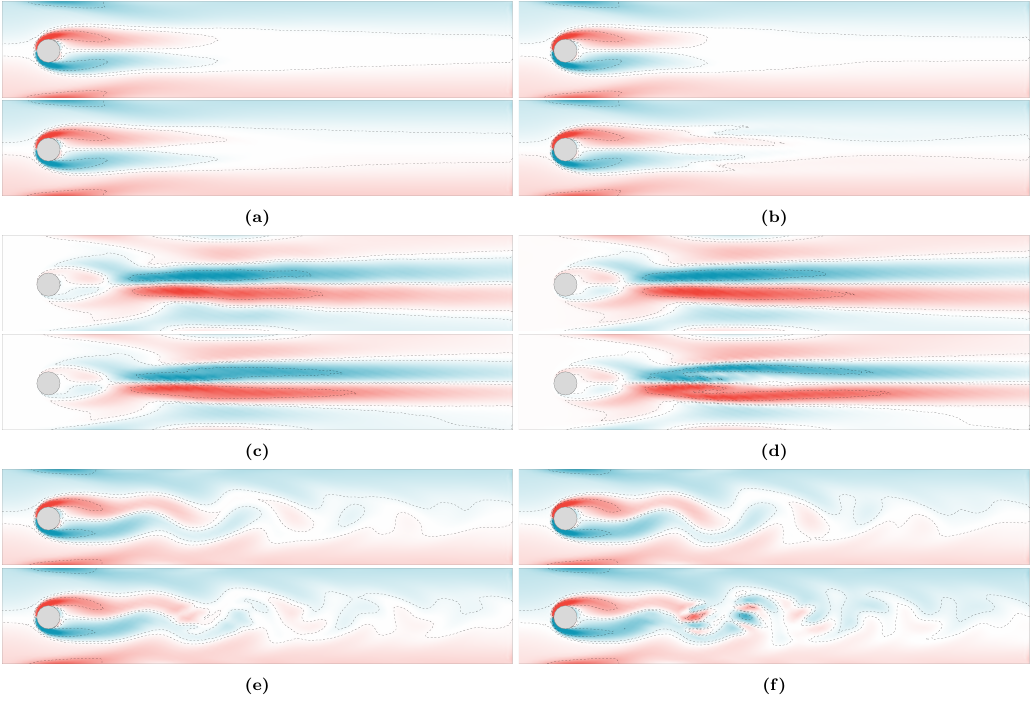}
    \caption{Vorticity of the FOM (top) and ROM (bottom) for different representative fields: mean flow, shift mode, and snapshot at the same time instance, for $Re=52$~\textbf{(a)-(c)-(e)} and $Re=54$~\textbf{(b)-(d)-(f)}.}
    \label{fig:mean_shift_and_snapshot}
\end{figure*}

A comparison between the ROM and the FOM solutions is now given in terms of flow fields, by considering the mean flow, the so-called \textit{shift-mode}~\cite{noack2003jfm}, and a snapshot of the limit cycle at one point in time.
Specifically, the shift-mode is here computed as the difference between the mean flow and the steady solution at a given value of the Reynolds number, and is therefore a way to visualise the transient. Note that, unlike~\cite{noack2003jfm}, this is only a reconstructed field, and not part of the ROM.
Fig.~\ref{fig:mean_shift_and_snapshot} compares the FOM flow fields with those computed with the ROM expanded around $Re_0 = 70$, for two different values of the Reynolds number: $Re = 52$ and $Re = 54$.
Despite the significant extrapolation in $Re$ due to the large gap between the expansion point and the evaluation points, the predictions remain closely aligned with the FOM results, with the deviation appearing at $Re = 54$ being confined to a small region of the domain.

The accuracy of the results presented so far is even more noteworthy when accounting for the computational advantage that the proposed ROM provides over the FOM. As reported in Table~\ref{tab:computational_cost}, the offline cost of building the ROM is significantly lower than that of a single FOM simulation, whilst the online cost is negligible. In particular, constructing the reduced model is $283$ times cheaper than computing the FOM until the flow is fully developed. Moreover, a single ROM can capture the system behaviour across a range of $Re$, while a new FOM simulation must be run for each value of the parameter. The reported times refer to computations run on a machine equipped with an Intel i9-12900K CPU and 32GB of RAM.

\begin{table}[thb]
    \centering
    \begin{tabular}{c|c|c}
         & FOM & ROM (offline) \\
         \hline
        Computational time & 7.55 hours & 96 seconds \\
        \# of linear systems solved & 30875 & 31
    \end{tabular}
    \caption{Comparison of computational cost between the construction of the ROM and a single FOM simulation at a given $Re$ with initial conditions around the steady solution, and up to the periodic steady state. Computational speed-up: $283$. Note that the ROM time is a one-off cost for a range of $Re$, while the FOM time refers to a single $Re$.}
    \label{tab:computational_cost}
\end{table}

In addition to reconstructing the various full fields, the method possesses the unique feature of providing an \textit{a priori} measure of the global error. In this contribution, we refer to an {\it a priori} measure as an error that do not need the computation of a full-order solution. It should be noted that this designation should not be confused with {\it a priori} errors that are based on a numerical analysis, and can be theoretically derived. Here, the a priori error is provided as a by-product of the reduction method itself. On the other hand, we will refer to an {\it a posteriori} error when a comparison to a full-order solution is needed.

Specifically, since the invariance property is enforced only up to a finite polynomial order, the ROM will solve Eq.~\eqref{eq:NS} up to a residual, which represents an unbalanced stress distribution.
Approximating the velocity and pressure fields $(\delta\bfu$, $\delta p)$ that would result from such distribution yields an estimate of the full field ROM error. In the present study, to quantify this error at any given $Re$, we seek the vector $\delta \bfy = [\delta \bfu, \delta p, \eta]^T$ which satisfies $\mathcal{A}(\delta \bfy) = \bfr$.
An estimator of the normalised root-mean-square error (NRMSE) of the velocity is then computed as the root-mean-square of $\delta\bfu$, normalised by the maximum inlet velocity.

\begin{figure}[ht!]
    \includegraphics[width=\linewidth]{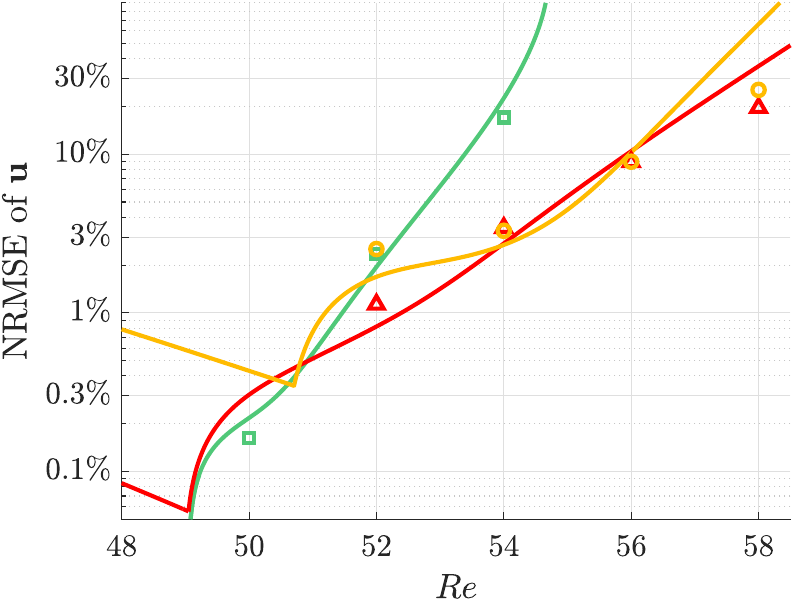}
    \caption{NRMSE of the velocity. \textit{A posteriori} evaluation and \textit{a priori} prediction for the ROM at $Re_0 = Re_{c}$ (\textcolor[HTML]{50c878}{$\square$, \rule[0.5ex]{0.2cm}{1pt}}) \cite{Carini2015}, $Re_0 = 70$ (\textcolor[HTML]{ff0000}{$\triangle$, \rule[0.5ex]{0.2cm}{1pt}}) and $Re_0 = 80$ (\textcolor[HTML]{ffbb00}{\scalebox{1.3}{$\circ$}, \rule[0.5ex]{0.2cm}{1pt}}); the a posteriori error is the NRMSE of the difference between the ROM and FOM velocity fields.}    
    \label{fig:normalised_RMSE}
\end{figure}

The quality of this estimator is assessed by comparing it with the NRMSE of the true error, evaluated \textit{a posteriori} as the difference between the FOM and ROM velocity fields.
As shown in Fig.~\ref{fig:normalised_RMSE}, the error predicted a priori closely matches the a posteriori evaluation. We note that our estimator differs from the one in \cite{Carini2015}, which evaluates the validity of the expansion without quantifying the error. As demonstrated in Appendix~\ref{sec:error_comparison}, our NRMSE prediction serves an equivalent purpose while providing a quantitative measure. This improvement allows for threshold-based selection rather than a binary assessment, offering finer control. It is also worth underlining that the {\it a posteriori} error reported in Fig.~\ref{fig:normalised_RMSE} is only given here for legitimating that the {\it a priori} error closely follows the same trend, which means that the validity limit of the ROM can be easily computed as a by-product of the method, without the need of full-order simulation.

\begin{figure}
    \includegraphics[width=\columnwidth]{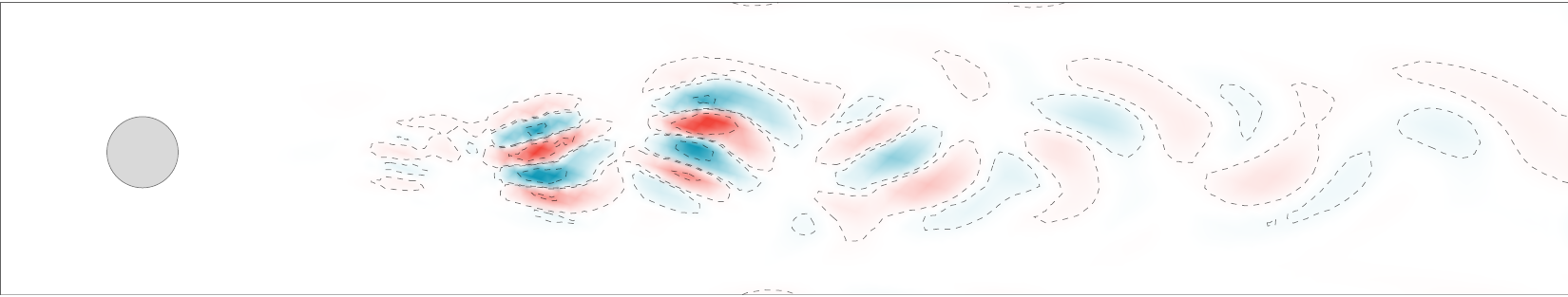}\\
    \includegraphics[width=\columnwidth]{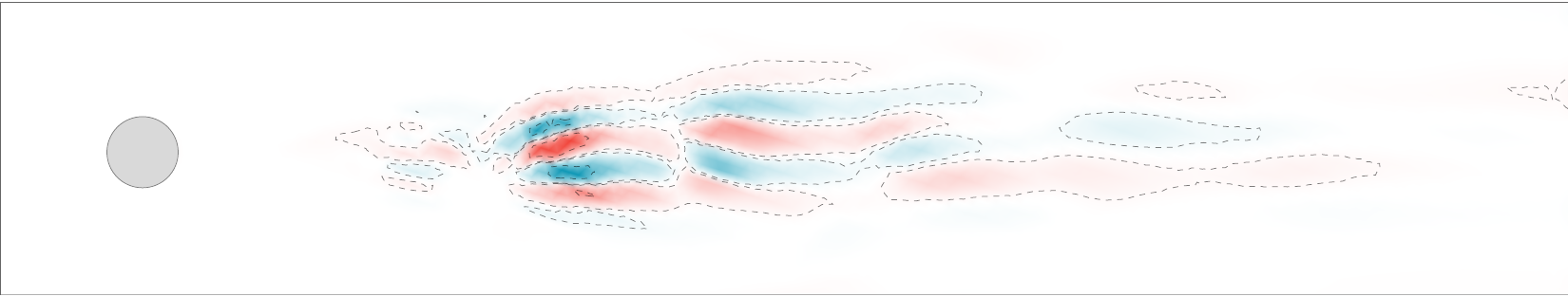}
    \caption{Full field vorticity error, normalised by maximum vorticity. \textit{A posteriori} evaluation and \textit{a priori} prediction for the $Re_0 = 70$ ROM evaluated at $Re = 54$; the a posteriori evaluation is the difference between the ROM and FOM vorticity fields. The values range from -10\% (\textcolor[RGB]{15,152,182}{\rule
    {0.2cm}{0.2cm}}) to +10\%
    (\textcolor[RGB]{241,68,58}{\rule{0.2cm}{0.2cm}}).}
    \label{fig:error_prediction_70@54}
\end{figure}

Further confirmation of the quality of the proposed error prediction is shown in Fig.~\ref{fig:error_prediction_70@54}.
Here, the vorticity of the true velocity error and that of $\delta\bfu$ are shown for the $Re_0 = 70$ ROM evaluated at $Re = 54$. Beyond accurately predicting the NRMSE, the estimate demonstrates a strong agreement with the error's spatial features and intensity.

\section{Conclusions}
\label{sec:conclusions}
An efficient reduced-order modelling procedure for flows undergoing Hopf bifurcations has been derived using the parametrisation method for invariant manifolds. This simulation-free method operates directly on the Navier-Stokes equations and only requires knowledge about the geometry, without relying on time-integrated snapshots.
A single computation at a fixed value of the bifurcating parameter produces a parametric ROM with minimal dimension: two coordinates to describe the oscillations and one to embed the parameter-dependence.
Nonetheless, the validity of this ROM spans a range of parameter values, allowing it to retrieve the pre-critical behaviour, with stable steady-state solutions, as well as the bifurcation point and the post-critical behaviour, characterised by limit cycle oscillations.

The method offers a systematic way to construct an invariant manifold and the associated embedded dynamics, ensuring that the ROM captures the most relevant features of the original system's behaviour. It also provides an \textit{a priori} measure of the error, which allows estimating the range of validity of the ROM in the absence of a reference solution.

The numerical results show good agreement between the ROM predictions and full-order model simulations, whilst providing a remarkable speed-up. As compared to the existing literature, it has been shown that the developments led in~\cite{Carini2015} used an equivalent framework, albeit not mentioning the parametrisation method, nor proposing other styles of parametrisation. By uncovering this link, this study places this development in a general framework and opens the door to more dedicated studies related to finding the best parameterisation, \textit{i.e.} one that would offer the largest validity range~\cite{Stoychev:failing}.

The expansion point at which the ROM is computed has also been varied in the present study, whereas~\cite{Carini2015} limited their numerical results to $Re_0 = Re_c$. By doing so, it has been shown that the ROM could predict the bifurcation point even when computed at a pre-critical parameter value. It has also been numerically demonstrated that the best solutions are found when parametrising the unstable manifold, when the ROM is computed at a post-critical value. Specifically, the range within which the method shows a good approximation to the limit cycle has been increased from 4\% to 10\%. The question of finding the best parametrisation styles has not been explored and is left for future works.

Further development of this work could be the application of the technique to other bifurcating flows, specifically considering sub-critical Hopf bifurcations or successive bifurcations. Another interesting outlook could be generalisation of the framework to flows interacting with deformable structures.

\section{Acknowledgements}
The authors are thankful to Fran\c{c}ois Gallaire for the insightful feedback on the manuscript.
Alessio Colombo acknowledges the support of ST Microelectronics under the framework of the Joint Research Platform with the Politecnico di Milano, STEAM.
Attilio Frangi acknowledges the PRIN 2022 Project “DIMIN- DIgital twins of nonlinear MIcrostructures with iNnovative model-order-reduction strategies” 
(No. 2022XATLT2) funded by the European Union - NextGenerationEU.

\bibliographystyle{unsrt}
\bibliography{biblioROM}

\begin{thebibliography}{10}

\bibitem{PODturbflows}
G.~Berkooz, P.~Holmes, and J.~L. Lumley.
\newblock The proper orthogonal decomposition in the analysis of turbulent flows.
\newblock {\em Annual Review of Fluid Mechanics}, 25:539--575, 1993.

\bibitem{noack2003jfm}
B.~R. Noack, K.~Afanasiev, M.~Morzy{\'n}ski, G.~Tadmor, and F.~Thiele.
\newblock A hierarchy of low-dimensional models for the transient and post-transient cylinder wake.
\newblock {\em Journal of Fluid Mechanics}, 497:335--363, 2003.

\bibitem{DMD2014}
J.~H. Tu, C.~W. Rowley, D.~M. Luchtenburg, S.~L. Brunton, and J.~N. Kutz.
\newblock On dynamic mode decomposition: Theory and applications.
\newblock {\em Journal of Computational Dynamics}, 1(2):391--421, 2014.

\bibitem{kutz2016book}
J.~N. Kutz, S.~L. Brunton, B.~W. Brunton, and J.~L. Proctor.
\newblock {\em Dynamic mode decomposition: data-driven modeling of complex systems}.
\newblock SIAM, 2016.

\bibitem{rowley2017arfm}
C.~W. Rowley and S.~Dawson.
\newblock Model reduction for flow analysis and control.
\newblock {\em Annual Review of Fluid Mechanics}, 49(1):387--417, 2017.

\bibitem{loiseau2018jfm}
J.-C. Loiseau, B.~R. Noack, and S.~L. Brunton.
\newblock Sparse reduced-order modelling: sensor-based dynamics to full-state estimation.
\newblock {\em Journal of Fluid Mechanics}, 844:459--490, 2018.

\bibitem{brunton2020arfm}
S.~L. Brunton, B.~R. Noack, and P.~Koumoutsakos.
\newblock Machine learning for fluid mechanics.
\newblock {\em Annual review of fluid mechanics}, 52(1):477--508, 2020.

\bibitem{kaszas2022prf}
B.~Kasz{\'a}s, M.~Cenedese, and G.~Haller.
\newblock Dynamics-based machine learning of transitions in {C}ouette flow.
\newblock {\em Physical Review Fluids}, 7(8):L082402, 2022.

\bibitem{bruntonSINDy}
S.~L. Brunton, J.~L. Proctor, and J.~N. Kutz.
\newblock Discovering governing equations from data by sparse identification of nonlinear dynamical systems.
\newblock {\em Proceedings of the National Academy of Sciences (PNAS)}, 113(15):3932--3937, 2016.

\bibitem{sipp2016amr}
D.~Sipp and P.~J. Schmid.
\newblock Linear closed-loop control of fluid instabilities and noise-induced perturbations: a review of approaches and tools.
\newblock {\em Applied Mechanics Reviews}, 68(2):020801, 2016.

\bibitem{deng2020jfm}
N.~Deng, B.~R. Noack, M.~Morzy{\'n}ski, and L.~Pastur.
\newblock Low-order model for successive bifurcations of the fluidic pinball.
\newblock {\em Journal of fluid mechanics}, 884:A37, 2020.

\bibitem{schmid2022annurev}
P.~J. Schmid.
\newblock Dynamic mode decomposition and its variants.
\newblock {\em Annual Review of Fluid Mechanics}, 54(Volume 54, 2022):225--254, 2022.

\bibitem{HaragusIooss}
M.~Haragus and G.~Iooss.
\newblock {\em Local bifurcations, center manifolds, and normal forms in infinite dimensional systems}.
\newblock EDP Science, 2009.

\bibitem{TITI1990}
E.~S. Titi.
\newblock On approximate inertial manifolds to the {N}avier-{S}tokes equations.
\newblock {\em Journal of Mathematical Analysis and Applications}, 149(2):540--557, 1990.

\bibitem{Carini2015}
M.~Carini, F.~Auteri, and F.~Giannetti.
\newblock Centre manifold reduction of bifurcating flows.
\newblock {\em Journal of Fluid Mechanics}, 767:109--145, 2015.

\bibitem{GallairePush}
F.~Gallaire, E.~Boujo, V.~Mantic-Lugo, C.~Arratia, B.~Thiria, and P.~Meliga.
\newblock Pushing amplitude equations far from threshold: application to the supercritical hopf bifurcation in the cylinder wake.
\newblock {\em Fluid Dynamics Research}, 48:061401, 2016.

\bibitem{Buza:NS}
G.~Buza.
\newblock Spectral submanifolds of the {N}avier-{S}tokes equation.
\newblock {\em SIAM Journal on Applied Dynamical Systems}, 23(2):1052--1089, 2024.

\bibitem{stuart1958}
J.~T. Stuart.
\newblock On the non-linear mechanics of hydrodynamic stability.
\newblock {\em Journal of Fluid Mechanics}, 4(1):1--21, 1958.

\bibitem{lusch2018nature}
B.~Lusch, J.~N. Kutz, and S.~L. Brunton.
\newblock Deep learning for universal linear embeddings of nonlinear dynamics.
\newblock {\em Nature communications}, 9(1):4950, 2018.

\bibitem{chen2018neuralODE}
R.~T.~Q. Chen, Y.~Rubanova, J.~Bettencourt, and D.~K. Duvenaud.
\newblock Neural ordinary differential equations.
\newblock {\em Advances in Neural Information Processing Systems (NeurIPS)}, 31:6571--6583, 2018.

\bibitem{ssmlearn}
M.~Cenedese, J.~Ax{\aa}s, B.~B{\"a}uerlein, K.~Avila, and G.~Haller.
\newblock Data-driven modeling and prediction of non-linearizable dynamics via spectral submanifolds.
\newblock {\em Nature Communications}, 13(1):872, 2022.

\bibitem{Amsallem2009}
D.~Amsallem and C.~Farhat.
\newblock Interpolation method for adapting reduced-order models and application to aeroelasticity.
\newblock {\em AIAA Journal}, 46(7):1803--1813, 2009.

\bibitem{Rozza2007RB}
G.~Rozza, D.~B.~P. Huynh, and A.~T. Patera.
\newblock Reduced basis approximation and a posteriori error estimation for parametrized partial differential equations.
\newblock {\em Archives of Computational Methods in Engineering}, 14(3):229--275, 2007.

\bibitem{Peherstorfer2016}
B.~Peherstorfer and K.~Willcox.
\newblock Data-driven operator inference for non-intrusive projection-based model reduction.
\newblock {\em Journal of Computational Physics}, 306:196--215, 2016.

\bibitem{Haken81}
A.~Wunderlin and H.~Haken.
\newblock Generalized {G}inzburg-{L}andau equations, slaving principle and center manifold theorem.
\newblock {\em Zeitschrift für Physik B Condensed Matter}, 44:135--141, 1981.

\bibitem{gucken83}
J.~Guckenheimer and P.~Holmes.
\newblock {\em Nonlinear oscillations, dynamical systems and bifurcations of vector fields}.
\newblock Springer-Verlag, New-York, 1983.

\bibitem{Cabre3}
X.~Cabr{\'e}, E.~Fontich, and R.~de~la Llave.
\newblock The parameterization method for invariant manifolds. {III}. {O}verview and applications.
\newblock {\em J. Differential Equations}, 218(2):444--515, 2005.

\bibitem{Haro}
A.~Haro, M.~Canadell, J.-L. Figueras, A.~Luque, and J.-M. Mondelo.
\newblock {\em The parameterization method for invariant manifolds. From rigorous results to effective computations}.
\newblock Springer, Switzerland, 2016.

\bibitem{Murdock}
J.~Murdock.
\newblock {\em Normal forms and unfoldings for local dynamical systems}.
\newblock Springer monographs in Mathematics, New-York, 2003.

\bibitem{Haller2016}
G.~Haller and S.~Ponsioen.
\newblock Nonlinear normal modes and spectral submanifolds: existence, uniqueness and use in model reduction.
\newblock {\em Nonlinear Dynamics}, 86(3):1493--1534, 2016.

\bibitem{JAIN2021How}
S.~Jain and G.~Haller.
\newblock How to compute invariant manifolds and their reduced dynamics in high-dimensional finite-element models.
\newblock {\em Nonlinear Dynamics}, 107:1417--1450, 2022.

\bibitem{ReviewROMGEOMNL}
C.~Touz{\'e}, A.~Vizzaccaro, and O.~Thomas.
\newblock Model order reduction methods for geometrically nonlinear structures: a review of nonlinear techniques.
\newblock {\em Nonlinear Dynamics}, 105:1141--1190, 2021.

\bibitem{vizza21high}
A.~Vizzaccaro, A.~Opreni, L.~Salles, A.~Frangi, and C.~Touz\'e.
\newblock High order direct parametrisation of invariant manifolds for model order reduction of finite element structures: application to large amplitude vibrations and uncovering of a folding point.
\newblock {\em Nonlinear Dynamics}, 110:525--571, 2022.

\bibitem{li2021periodic}
M.~Li, S.~Jain, and G.~Haller.
\newblock Nonlinear analysis of forced mechanical systems with internal resonance using spectral submanifolds -- part {I}: {P}eriodic response and forced response curve.
\newblock {\em Nonlinear Dynamics}, 110:1005--1043, 2022.

\bibitem{opreni22high}
A.~Opreni, A.~Vizzaccaro, C.~Touz\'e, and A.~Frangi.
\newblock High order direct parametrisation of invariant manifolds for model order reduction of finite element structures: application to generic forcing terms and parametrically excited systems.
\newblock {\em Nonlinear Dynamics}, 111:5401--5447, 2023.

\bibitem{vizza2023superharm}
A.~Vizzaccaro, G.~Gobat, A.~Frangi, and C.~Touz{\'e}.
\newblock Direct parametrisation of invariant manifolds for non-autonomous forced systems including superharmonic resonances.
\newblock {\em Nonlinear Dynamics}, 112:6255--6290, 2024.

\bibitem{Martin:rotation}
A.~Martin, A.~Opreni, A.~Vizzaccaro, M.~Debeurre, L.~Salles, A.~Frangi, O.~Thomas, and C.~Touz{\'e}.
\newblock Reduced order modeling of geometrically nonlinear rotating structures using the direct parametrisation of invariant manifolds.
\newblock {\em Journal of Theoretical, Computational and Applied Mechanics}, 10430, 2023.

\bibitem{Frangi:electromech}
A.~Frangi, A.~Colombo, A.~Vizzaccaro, and C.~Touzé.
\newblock Reduced order modelling of fully coupled electro-mechanical systems through invariant manifolds with applications to microstructures.
\newblock {\em International Journal for Numerical Methods in Engineering}, 126(3):e7641, 2025.

\bibitem{Pinho:shells}
F.~A.~X.~Carneiro Pinho, M.~Amabili, Z.~J.~G.~N.~Del Prado, and F.~M.~Alves da~Silva.
\newblock Nonlinear forced vibration analysis of doubly curved shells via the parameterization method for invariant manifold.
\newblock {\em Nonlinear Dynamics}, 112:20677–20701, 2024.

\bibitem{vdb:centerMORE}
J.~B. {van den Berg}, W.~Hetebrij, and B.~Rink.
\newblock More on the parameterization method for center manifolds.
\newblock 2020.

\bibitem{MingwuLi2024}
M.~Li and L.~Wang.
\newblock Parametric model reduction for a cantilevered pipe conveying fluid via parameter-dependent center and unstable manifolds.
\newblock {\em International Journal of Non-Linear Mechanics}, 160:104629, 2024.

\bibitem{Stabile:follow}
A.~de~Figueiredo~Stabile, A.~Vizzaccaro, L.~Salles, A.~Colombo, A.~Frangi, and C.~Touzé.
\newblock Reduced-order modelling of parameter-dependent systems with invariant manifolds: application to hopf bifurcations in follower force problems.
\newblock {\em International Journal of Nonlinear Mechanics}, 177:105133, 2025.

\bibitem{Cabre1}
X.~Cabr{\'e}, E.~Fontich, and R.~de~la Llave.
\newblock The parameterization method for invariant manifolds. {I}. {M}anifolds associated to non-resonant subspaces.
\newblock {\em Indiana Univ. Math. J.}, 52(2):283--328, 2003.

\bibitem{cabre2}
X.~Cabr{\'e}, E.~Fontich, and R.~de~la Llave.
\newblock The parameterization method for invariant manifolds. {II}. {R}egularity with respect to parameters.
\newblock {\em Indiana Univ. Math. J.}, 52(2):329--360, 2003.

\bibitem{vdb:center2020}
J.~Bouwe {van den Berg}, W.~Hetebrij, and B.~Rink.
\newblock The parameterization method for center manifolds.
\newblock {\em Journal of Differential Equations}, 269(3):2132--2184, 2020.

\bibitem{artDNF2020}
A.~Vizzaccaro, Y.~Shen, L.~Salles, J.~Blahos, and C.~Touz{\'e}.
\newblock Direct computation of nonlinear mapping via normal form for reduced-order models of finite element nonlinear structures.
\newblock {\em Computer Methods in Applied Mechanics and Engineering}, 284:113957, 2021.

\bibitem{Stoychev:failing}
A.K. Stoychev and U.J. R{\"o}mer.
\newblock Failing parametrizations: what can go wrong when approximating spectral submanifolds.
\newblock {\em Nonlinear Dynamics}, 111:5963--6000, 2023.

\bibitem{zdravkovich1997}
M.~M.~M. Zdravkovich.
\newblock {\em Flow around circular cylinders: Volume 1: Fundamentals}, volume~1.
\newblock Oxford university press, 1997.

\bibitem{TouzeCISM}
C.~Touz{\'e}.
\newblock Normal form theory and nonlinear normal modes: theoretical settings and applications.
\newblock In G.~Kerschen, editor, {\em Modal Analysis of nonlinear Mechanical Systems}, pages 75--160, New York, NY, 2014. Springer Series CISM courses and lectures, vol. 555.

\bibitem{TouzeCISM2}
C.~Touz{\'e} and A.~Vizzaccaro.
\newblock Nonlinear normal modes as invariant manifolds for model order reduction.
\newblock In C.~Touz{\'e} and A.~Frangi, editors, {\em Model Order Reduction for Design, Analysis and Control of Nonlinear Vibratory Systems}, pages 59--116, New York, NY, 2024. Springer Series CISM courses and lectures, vol. 614.

\bibitem{SIPP_LEBEDEV_2007}
D.~Sipp and A.~Lebedev.
\newblock Global stability of base and mean flows: a general approach and its applications to cylinder and open cavity flows.
\newblock {\em Journal of Fluid Mechanics}, 593:333–358, 2007.

\bibitem{Guillot:recast}
L.~Guillot, B.~Cochelin, and C.~Vergez.
\newblock A generic and efficient {T}aylor series-based continuation method using a quadratic recast of smooth nonlinear systems.
\newblock {\em International Journal for Numerical Methods in Engineering}, 119(4):261--280, 2019.

\bibitem{Schafer1996}
M.~Sch{\"a}fer, S.~Turek, F.~Durst, E.~Krause, and R.~Rannacher.
\newblock {\em Benchmark Computations of Laminar Flow Around a Cylinder}, pages 547--566.
\newblock Vieweg+Teubner Verlag, Wiesbaden, 1996.

\bibitem{Stabile:morfesym}
A.~de~Figueiredo~Stabile, C.~Touzé, and A.~Vizzaccaro.
\newblock Normal form analysis of nonlinear oscillator equations with automated arbitrary order expansions.
\newblock {\em Journal of Theoretical, Computational and Applied Mechanics}, 13234, 2025.

\bibitem{GROLET:high25}
A.~Grolet, A.~Vizzaccaro, M.~Debeurre, and O.~Thomas.
\newblock High order invariant manifold model reduction for systems with non-polynomial non-linearities: Geometrically exact finite element structures and validity limit.
\newblock {\em International Journal of Non-Linear Mechanics}, 178:105138, 2025.

\end{thebibliography}

\appendix

\clearpage
\section{Discrete formulation and numerical treatment of the homological equations}
\label{sec:discrete}

In this section, a discrete counterpart to the continuous formulation discussed in Section~\ref{sec:detailed_method} is presented. We focus here specifically on the discretised problem and its numerical treatment, as well as give more thorough details on the quantities to be computed. Since the algorithmic procedure is analogous, the presentation follows closely the framework established in \cite{vizza2023superharm}, though applied to a different context.\\
The discrete approximations of the continuum operators introduced in Section~\ref{sec:detailed_method} will be denoted here by bold capital letters ($\bfm{A}, \bfm{B}, \bfm{Q}$), representing their spatially discretised counterparts. These operators, originally defined in Eq.~\eqref{eq:operators}, are implicitly constructed through an appropriate numerical method (e.g., Finite Elements) while preserving the same functional relationships as their continuous counterparts. The same convention extends to all the discretised fields.
In this context, the set of order 1 homological equations reads:
\begin{equation}
    \bfm{B} \bfLambda \bfPhi = \bfm{A} \bfPhi,
\end{equation}
where the same structure as the eigenvalue problem can once again be seen. 
At order $p\geq2$, the homological equation reads:
\begin{equation}\label{eq:invorderp}
    \bfm{B} [\nabla_\bfz \bfm{W}(\bfz)\bff(\bfz)]_p = \bfm{A}[\bfm{W}(\bfz)]_p + \bfm{Q}[\bfm{W}(\bfz),\bfm{W}(\bfz)]_p.
\end{equation}
In Eq.~\eqref{eq:invorderp}, the shortcut notation $[\, \cdot \,]_p$ represents an operator that selects only the terms with degree $p$.
In order to write the homological equation for a generic monomial $(p,k)$, the three terms are first studied separately.
On the right-hand side, one has:
\begin{subequations}
\label{eq:rhs_homological}
    \begin{align}
        \bfm{A}[\bfm{W} (\bfz)]_p &= \sum_{k=1}^{m_p} \bfm{A}\bfm{W}^{(p,k)} \bfz^{\bfalpha(p,k)}, \\
        [\bfm{Q} (\bfm{W}(\bfz),\bfm{W}(\bfz))]_p &= \sum_{k=1}^{m_p} \bfm{Q}^{(p,k)} \bfz^{\bfalpha(p,k)}, 
    \end{align}
\end{subequations}
where the quadratic terms are constructed from the product of lower-order terms. In particular, 
by exploiting the general relationship:
\begin{align*}
    \bfm{Q}(\bfm{W}^{(p_1,k_1)} &\bfz^{\bfalpha (p_1,k_1)} \, , \,   \bfm{W}^{(p_2,k_2)} \bfz^{\bfalpha (p_2,k_2)}) =\\ 
 &= \bfm{Q}(\bfm{W}^{(p_1,k_1)} ,\bfm{W}^{(p_2,k_2)})  \bfz^{\bfalpha (p_1,k_1) + \bfalpha (p_2,k_2)},
\end{align*}
for generic orders $p_1$ and $p_2$, the terms $\bfm{Q}^{(p,k)}$ at a given order $p$ can be computed in parallel:
\begin{align}
    \bfm{Q}^{(p,k)} & = 
    \sum_{p_1=1}^{p-1} 
    \sum_{k_1,k_2=1}^{m_{p_1},m_{p_2}} 
    \bfm{Q} (\bfm{W}^{(p_1,k_1)} ,\bfm{W}^{(p_2,k_2)}),  
    \\     
    & p_2: \quad p_2=p-p_1,
    \nonumber 
    \\
    & k : \quad \bfalpha(p,k)=\bfalpha (p_1,k_1) + \bfalpha (p_2,k_2).
    \nonumber
\end{align}
On the left-hand side, three contributions of order $p$ appear:
\begin{align}
    [\nabla_{\bfz} \bfm{W} (\bfz) \bff (\bfz)]_p &= \bfm{N}_1(\bfz) + \bfm{N}_2(\bfz) + \bfm{N}_3(\bfz),\\
    \bfm{N}_1(\bfz) &= \sum_{s=1}^{|\bfz|}  \bfm{W}^{(1,s)} [f_s(\bfz)]_p, \nonumber \\
    \bfm{N}_2(\bfz) &= \sum_{s=1}^{|\bfz|} \left( \sum_{j=1}^{|\bfz|} f_s^{(1,j)} z_j  \right) \frac{\partial [\bfm{W}(\bfz)]_p}{\partial z_s}, \nonumber \\
    \bfm{N}_3(\bfz) &= \sum_{s=1}^{|\bfz|} \left[ \frac{\partial [\bfm{W}(\bfz)]_{\dblnk} } {\partial z_s} [f_s(\bfz)]_{\dblnk} \right]_p, \nonumber
\end{align}
where the symbol $\dblnk$ has been introduced to indicate terms of order strictly larger than 1 and smaller than $p$.\\
Concerning the first term, one has:
\begin{equation}
    \bfm{N}_1(\bfz) = \sum_{s=1}^{|\bfz|}  \bfm{W}^{(1,s)} [f_s(\bfz)]_p = \sum_{k=1}^{m_p} \sum_{s=1}^{|\bfz|}  \bfPhi_s f_s^{(p,k)}\bfz^{\bfalpha(p,k)},
\end{equation}
where the unknown coefficients of the reduced dynamics appear.
In the second term, the choice $\bfLambda = [\bff^{(1,1)},\ldots,\bff^{(1,|\bfz|)}]$ (Eq.~\eqref{eq:Lambda}) implies:
\begin{equation}
     \sum_{j=1}^{|\bfz|} f_s^{(1,j)} z_j = \lambda_s z_s,
\end{equation}
so that one can write:
\begin{align}
    \bfm{N}_2(\bfz) &= \sum_{k=1}^{m_p} \sum_{s=1}^{|\bfz|} \lambda_s z_s \frac{\partial [\bfm{W}(\bfz)]_p}{\partial z_s} =\nonumber\\
    &= \sum_{k=1}^{m_p} \sum_{s=1}^{|\bfz|} \lambda_s \alpha_s(p,k) \bfm{W}^{(p,k)} \bfz^{\bfalpha(p,k)},
\end{align}
where the unknown coefficients of the mapping appear instead. By expressing this in the same form as Eqs.~\eqref{eq:rhs_homological} and using $\sigma^{(p,k)}$ as defined in the main text, this leads to:
\begin{align}
    \bfm{N}_2(\bfz) &= \sum_{k=1}^{m_p} \bfm{N}_2^{(p,k)} \bfz^{\bfalpha(p,k)},\\
    \bfm{N}_2^{(p,k)} &= \sigma^{(p,k)} \bfm{W}^{(p,k)}.
\end{align}
The third term only contains terms of lower order: no unknown is involved in its computation, meaning it contributes to the right-hand side of the homological equation. With an analogous expansion:
\begin{equation}
    \bfm{N}_3(\bfz) = \sum_{p=1}^{m_p} \bfm{N}_3^{(p,k)} \bfz^{\bfalpha(p,k)}.
\end{equation}
The total contribution to the $(p,k)$-th coefficients can then be expressed as:
\begin{align}
    \bfm{N}_3^{(p,k)} & = 
    \sum_{s=1}^{|\bfz|} 
    \sum_{p_W=2}^{p-1}
    \sum_{k_W,k_f=1}^{m_{p_W},m_{p_f}} 
    \alpha_s (p_W,k_W) \bfm{W}^{(p_W,k_W)} f_s^{(p_f,k_f)},
    \\     
    & p_f: \quad p_f=p+1-p_W ,
    \nonumber 
    \\
    & k: \quad \bfalpha(p,k)=\bfalpha (p_W,k_W) + \bfalpha (p_f,k_f) - \bfm{e}_s,
    \nonumber
\end{align}
with $\bfm{e}_s$ being the $s$-th vector of the canonical basis.
Denoting with $\bfm{R}^{(p,k)}$ the right-hand side term gathering all the known quantities:
\begin{equation}
    \bfm{R}^{(p,k)} = \bfm{Q}^{(p,k)} - \bfm{B} \left(\bfm{N}_3^{(p,k)} + \bfm{N}_2^{(p,k)} \right),   
\end{equation}
the homological equation can finally be written for an arbitrary monomial:
\begin{equation}
\label{eq:homological_discrete}
    \left( \sigma^{(p,k)} \bfm{B} - \bfm{A} \right) \bfm{W}^{(p,k)} + \sum_{s=1}^d \bfm{B} \bfPhi_s f_s^{(p,k)} = \bfm{R}^{(p,k)}.
\end{equation}
Since the equation is underdetermined, additional constraints must be enforced to ensure the solvability of this equation, as discussed in the main text. For illustrative purposes, we report here the full linear system in the case of normal form style, composed of Eq.~\eqref{eq:homological_discrete} endowed with the discretised Eqs.~\eqref{eq:constraints_normal_form_style}
The interested reader can find more detail on the procedure and the choices to be set for other styles, in the context of a direct solution, in~\cite{vizza2023superharm}, where a bordering technique is used as a key to operate from the physical space. When the normal form style is selected, the augmented system to be solved reads:
\begin{equation}
    \begin{bmatrix}
    \sigma^{(p,k)} \bfm{B}- \bfm{A} &   \bfm{B} \bfPhi_{\mathcal{R}} & \bfm{0}
    \\
    \bfPsi_{\mathcal{R}}^\star\bfm{B} &  \bfm{0}  & \bfm{0} \\
    \bfm{0}    &   \bfm{0}      & \bfm{I}
    \end{bmatrix}
   \begin{bmatrix}   \bfm{W}^{(p,k)}\\
   \bff^{(p,k)}_{\mathcal{R}}\\
   \bff^{(p,k)}_{\cancel{\mathcal{R}}}
    \end{bmatrix}
    =
    \begin{bmatrix}
    \bfm{R}^{(p,k)}\\
    \bfm{0} \\
    \bfm{0}
    \end{bmatrix},
\end{equation}
where the subscript $\mathcal{R}$ (resp. $\cancel{\mathcal{R}}$) indicates the collection of all the terms which are (resp. are not) resonant with the $(p,k)$-th monomial, as defined in Eq.~\eqref{eq:resonant_set}. Interestingly, the bordering technique was also proposed in~\cite{Carini2015} to operate from the physical degrees-of-freedom, but only the normal form style was made explicit, and the infinity of other possible solutions, including the graph style, were not commented on. Other choices are also possible to propose a direct procedure, as exemplified in~\cite{JAIN2021How}, where a norm-minimising procedure is adopted.


\section{Spectral analysis of the augmented Navier-Stokes system}\label{appendix:C}


In this appendix, the eigenvalue problem for the Studied Turek-Sch{\"a}fer benchmark is detailed. The eigenproblem at $Re_0$ reads:
\begin{equation}
\label{eq:NS-eig}
\begin{aligned}
&
\lambda \bfv_{\bfu}(\bfx) + 
\mathcal{L}_0(\bfv_{\bfu}(\bfx))
= 
v_{\eta} \triangle \bfu_0(\bfx) -
\nabla v_{p}(\bfx),
\\
&
 \nabla \cdot \bfv_{\bfu}(\bfx) = 0,
\\
&
\lambda v_{\eta} = 0.
\end{aligned}
\end{equation}
where $(\bfv_{\bfu}(\bfx), v_p(\bfx), v_\eta)$, and $\lambda$ are the generic eigenmode and eigenvalue.

\begin{figure}[hb]
    \includegraphics[width=\linewidth]{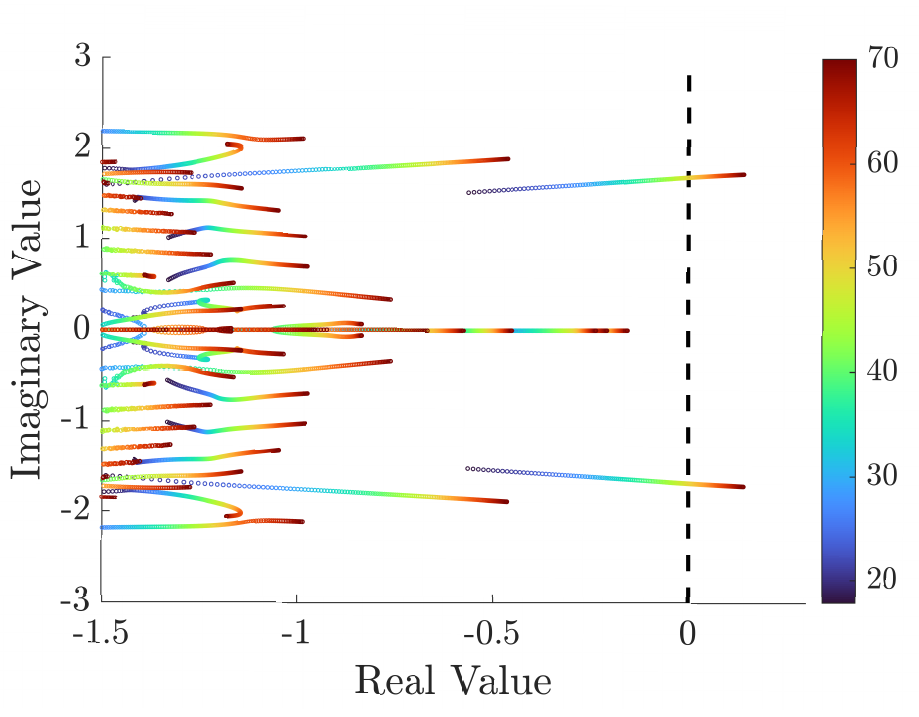}
    \caption{Eigenvalues trajectory for the flow past a cylinder, for Reynolds numbers varying from 18 to 70. A single pair of complex conjugates eigenvalues crosses the imaginary axis at $Re = Re_c$.}
    \label{fig:trajvp}
\end{figure}

In this eigenproblem, one has to distinguish the classical modes, stemming from the original Navier-Stokes problem, from a further mode arising from the addition of Eq.~\eqref{eq:NS-0c}, which is here referred to as the \textit{parameter mode}. The classical modes simply correspond to the case $v_{\eta}=0$, for which the original Navier-Stokes eigenproblem is retrieved:
\begin{equation*}
\begin{aligned}
&
\lambda \bfv_{\bfu}(\bfx) + 
\mathcal{L}_0(\bfv_{\bfu}(\bfx))
= -
\nabla v_{p}(\bfx),
\\
&
\nabla \cdot \bfv_{\bfu}(\bfx) = 0.
\end{aligned}
\end{equation*}
The eigenvalues of this problem are illustrated in Fig.~\ref{fig:trajvp} as $Re_0$ varies. A complex conjugate pair of eigenvalues crosses the imaginary axis at the Hopf bifurcation point $Re_c$, which in the main text has been simply denoted as $(\lambda,\bar{\lambda})$. These modes are the primary drivers of the slow dynamics and therefore must be included in $\bfphi$.

Regarding the above-mentioned parameter mode, the only solution to Eqs.~\eqref{eq:NS-eig} with a non-zero $v_\eta$ corresponds to the eigenvalue $\lambda=0$. This is also a slow mode that needs to be included in $\bfphi$. It is computed from:
\begin{equation*}
\begin{aligned}
&
\mathcal{L}_0(\bfv_{\bfu}(\bfx))
= 
\triangle \bfu_0(\bfx) -
\nabla v_{p}(\bfx),
\\
&
\nabla \cdot \bfv_{\bfu}(\bfx) = 0,
\end{aligned}
\end{equation*}
where $v_\eta$ has been arbitrarily set to 1.

The two bifurcating eigenmodes with eigenvalues $(\lambda,\bar{\lambda})$ and the parameter mode with eigenvalue $0$ represent the smallest possible subset of eigenmodes required to capture the slow dynamics of the full model. In fact, the first two are the main driver of the vortex shedding, while the latter is what enables the parameter-dependence in the ROM. The solution manifold is then parametrised by three coordinates, the first two being tangent to the modal coordinates of the bifurcating eigenmodes and the third being equal to the parameter variable $\eta'$:
\begin{equation}
    \bfz = 
    \begin{bmatrix}
        z_1\\
        z_2\\
        z_3
    \end{bmatrix}
    =
    \begin{bmatrix}
        \rho e^{+i\theta}\\
        \rho e^{-i\theta}\\
        \frac{1}{Re}-\frac{1}{Re_0}
    \end{bmatrix}.
\end{equation}


\section{Comparative analysis of ROM convergence validation methods}
\label{sec:error_comparison}
This appendix provides a formal comparison between the validity criterion of \cite{Carini2015} and our proposed {\it a priori}  NRMSE-based error estimator. We first reconstruct the logic of the original method, then demonstrate how our approach replicates its validity assessment while extending it with error quantification. The analysis shows that our framework strictly generalises the binary criterion while enabling application-dependent precision control through adjustable thresholds.

The original validation approach examines the convergence properties of amplitude expansions by analysing their asymptotic solutions. In particular, for a given value of the bifurcation parameter $Re$, the method identifies the limit cycle amplitude $\rho_{\infty}$ at successive expansion orders by solving the steady state condition $\partial_t \rho = 0$ in Eq.~\eqref{eq:ROM1}, which reduces to finding roots of the resulting polynomial. The ROM's validity is established when $\rho_{\infty}$ stabilises with increasing order refinement, confirming series convergence.

\begin{figure}[ht!]
    \includegraphics[width=\linewidth]{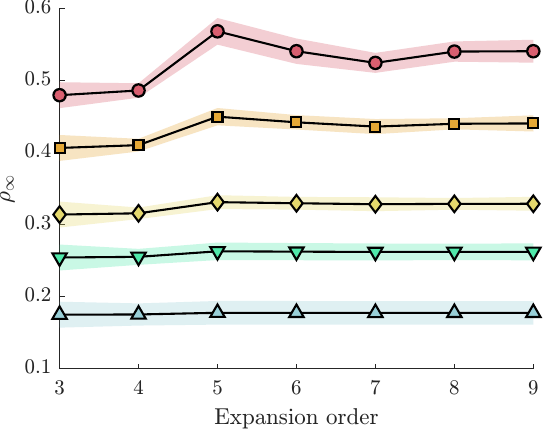}
    \caption{Convergence criteria comparison: limit cycle amplitude ({\color[HTML]{888888}$\blacktriangle,\blacktriangledown,\blacklozenge,\blacksquare,\bullet$}), proposed in \cite{Carini2015}, vs NRMSE prediction ({\color[HTML]{cccccc}\rule{2ex}{1.5ex}}) for the ROM at $Re_0 = Re_{c}$. The error bands are normalised to start with the same width, for visualisation purposes. Evaluation performed at $Re_0 = 49.5$ ({\color[HTML]{98cfd7}$\blacktriangle$\llap{\color{black}$\btriangle$}}, \textcolor[HTML]{c5eaf0}{\rule{2ex}{1.5ex}}), $Re_0 = 50.0$ ({\color[HTML]{4fe9aa}$\blacktriangledown$\llap{\color{black}$\triangledown$}}, \textcolor[HTML]{4fe9aa}{\rule{2ex}{1.5ex}}), $Re_0 = 50.5$ ({\color[HTML]{e5d96d}$\blacklozenge$\llap{\color{black}$\lozenge$}}, \textcolor[HTML]{e5d96d}{\rule{2ex}{1.5ex}}), $Re_0 = 51.5$ (\scalebox{0.8}{{\color[HTML]{e5a836}$\blacksquare$\llap{\color{black}$\square$}}}, \textcolor[HTML]{e5a836}{\rule{2ex}{1.5ex}}), $Re_0 = 52.5$ ({\color[HTML]{d95f70}$\bullet$\llap{\color{black}$\circ$}}, \textcolor[HTML]{d95f70}{\rule{2ex}{1.5ex}}).}    
    \label{fig:comparison_Carini_bifurcation}
\end{figure}

To establish equivalence between the methods, we evaluate the NRMSE predictor across increasing orders for corresponding $Re$ values. Convergence is demonstrated when the error estimate stabilises with order refinement, mirroring the $\rho_{\infty}$ stabilisation criterion.
Figure~\ref{fig:comparison_Carini_bifurcation} compares both approaches up to order $o=9$ for $Re \in \{49.5, \, 50.0, \, 50.5, \, 51.5, \, 52.5 \}$, using the ROM at $Re_0 = Re_{c}$ for consistent comparison. For direct comparability, the predicted NRMSE is visualised as bands around the $\rho_{\infty}$ trends, with width proportional to the actual estimated NRMSE. The two methods show strong agreement in assessing ROM validity: for $Re \leq 50.5$, both $\rho_{\infty}$ and the NRMSE plateau at $o \geq 7$, whereas at $Re = 51.5$ slight variability persists, suggesting proximity to the convergence limit. At $Re=52.5$, both exhibit significant order-to-order fluctuations, confirming non-convergence. Notably, however, the NRMSE remains modest ($\approx 1\%$) even in non-convergent regime, suggesting the ROM may still be usable in applications where such error levels are permissible. This underscores a key advantage of the NRMSE approach: while maintaining consistency with the binary criterion in detecting convergence, it provides additional flexibility in assessing tolerable error margins. We note that an analogous analysis for the model computed at $Re_0 = 70$ is precluded, as this would entail evaluating $\rho_{\infty}$ at an $\mathcal{O}(1)$ distance from the expansion point. However, similar convergence behaviour would be observed for transient amplitudes near the expansion point.

\end{document}